\documentclass[12pt]{article}
\usepackage{axodraw}
\usepackage{epsfig}
\usepackage{array}
\usepackage{float}
\usepackage{amstext}
\usepackage{rotating}
\usepackage{a4}
\usepackage{a4wide}
\usepackage{cite}

\newcommand{\be}{\begin{eqnarray}}
\newcommand{\ee}{\end{eqnarray}}

\newcommand{\ma}{\Delta m^2_{31}}

\def\nn{\nonumber}

\begin{document}

\title{
\Large 
Spontaneous R-parity violating type III seesaw}
\author{
Sandhya Choubey\thanks{email: \tt sandhya@hri.res.in},~~~
Manimala Mitra\thanks{email: \tt mmitra@hri.res.in}
\\\\
{\normalsize \it Harish--Chandra Research Institute,}\\
{\normalsize \it Chhatnag Road, Jhunsi, 211019 Allahabad, India }\\ \\ 
}
\date{ \today}
\maketitle
\vspace{-0.8cm}
\begin{abstract}
\noindent  

We present a model where neutrino masses are generated 
by a combination of spontaneous R-parity violation 
and Type III seesaw. 
In addition to the 
usual MSSM particle content, our model 
consists of one extra triplet matter chiral superfield  containing  
heavy SU(2) triplet fermions and its superpartners. R-parity is broken 
spontaneously when the sneutrinos associated with the 
one heavy neutrino as well as the three light neutrinos 
get vacuum expectation values, giving rise to the 
 mixed $8\times 8$ neutralino-neutrino mass matrix. 
We show that our model can comfortably explain all the existing 
neutrino oscillation data. Due to the presence of the 
triplet fermion, we have a pair of 
additional heavy charged leptons which mix with the 
standard model charged leptons and the charginos. 
This gives rise to a $6\times 6$ chargino-charged lepton 
mass matrix, with 6 massive eigenstates. Finally we discuss
about the different R-parity violating possible decay modes and 
the distinctive collider signatures which our model offers.

\end{abstract}

\newpage

\section{Introduction}

Despite its inimitable success, it is now universally accepted that 
the standard model of particle physics is only the low energy 
limit of a more complete theory of elementary particles. 
Various  experimental evidences on the existence 
of neutrino flavor oscillations have proved beyond doubt that the 
standard model needs to extended in order to explain neutrino masses and 
flavor mixing.  
Experiments like SNO, KamLAND, K2K and MINOS 
\cite{solar, kl, k2k, minos} provide information on the two
mass square differences $\Delta m_{21}^2$ and $\Delta m_{31}^2$ 
\footnote{We define $\Delta m_{ij}^2=m_i^2-m_j^2$.} and on the two
mixing angles $\theta_{12}$ and $\theta_{23}$. The third mixing angle
$\theta_{13}$ is not yet determined, but from the null result of the CHOOZ \cite{chooz}experiment it 
 is known to be certainly small. The
current $3 \sigma$  allowed intervals  of the oscillation parameters are
given as \cite{limits}
\be
 7.1 \times 10^{-5} \rm{eV^2}<\Delta m_{21}^2 < 8.3\times 10^{-5}
 \rm{eV^2},\hspace*{0.1cm} 2.0 \times 10^{-3} \rm{eV^2}<\Delta m_{31}^2 <
 2.8\times 10^{-3} \rm{eV^2}
\label{eq:delmasses}
\ee
 \be
   0.26< \sin^2 \theta_{12}<0.42,\hspace*{0.1cm}    0.34 <\sin^2
   \theta_{23}<0.67 , \hspace*{0.1cm}  \sin^2 \theta_{13}<0.05 ~.
\label{eq:angles}
\ee 
Any theory of physics beyond the standard model must 
offer a natural explanation of the neutrino masses. The two main 
challenges involved at this frontier are, (i) 
explaining the smallness of neutrinos 
masses, which are at least 12 orders of magnitude smaller compared to the 
top mass, and (ii) explaining 
the mixing pattern which is very distinct from the 
mixing observed in the quark sector. Very small Majorana 
neutrino masses can be generated by the 
dimension 5 operator $\frac{1}{\Lambda}LLHH$ \cite{dim5}, 
where the masses are suppressed naturally by the scale of new 
physics $\Lambda$. Note that this term breaks lepton number 
 which is mandatory for the generation of 
Majorana masses. In the seesaw mechanism \cite{seesaw}, 
$\Lambda$ is associated with the mass of new heavy particle(s). 
In the so-called Type I seesaw \cite{seesaw}, the new particles are 
heavy right-handed neutrinos which are singlets under the 
standard model gauge group. In Type III seesaw 
\cite{type3,type3all,Abada:2007ux, lfv,collidertype3, 
collidertypeall,type3others,type3us}, 
the new particles are heavy triplet fermions with hypercharge $Y=0$.  
The  fermions therefore are self-conjugate and 
belong to the adjoint representation of SU(2). Small Majorana 
masses can also be realized in the Type II seesaw models
\cite{type2}, where the model is extended by including one 
heavy complex scalar with hypercharge 
$Y=2$, which transforms as a triplet of SU(2). 
Observed neutrino mixing can be obtained very naturally by imposing
flavor symmetry. Among the numerous viable flavor symmetry models 
\cite{flvgen}, the models based on the group $A_4$ are  the most popular ones
\cite{flv}.

Among the  main drawbacks of the standard model is the problem 
of explaining the stability of the Higgs mass. Supersymmetry 
in particle physics \cite{susy}
has been one of the most widely accepted way of 
alleviating this problem. Supersymmetry demands that for every fermion/boson 
in the model, there is a superpartner which is a boson/fermion. 
In the limit of exact supersymmetry, this results in the cancellation 
of all quadratic divergences of the Higgs mass, which is hence 
naturally stabilized against quantum corrections. The most general 
superpotential allows for terms which break lepton number and 
baryon number\footnote{All possible terms in the 
Minimal Supersymmetric Standard Model (MSSM) superpotential 
are discussed in section 2 and given in Eqs. (\ref{eq:wmssm}) 
and (\ref{eq:rpv}).}.
However these lepton and baryon number violating couplings are 
severely constrained
by non-observation of proton decay  and data on heavy flavor physics from 
Belle and Babar \cite{bfact}. In order to avoid all such terms in the 
superpotential, one imposes a $Z_2$ symmetry called R-parity, 
which is defined as $R_p = (-1)^{3(B-L)+2S}$ 
\cite{Rparityold,Rparityold2,Rparity}, where 
$B$ and $L$ are the baryon and lepton number of the particle  
and $S$ is the spin. In the limit where R-parity is conserved 
both lepton and baryon number are conserved, whereas the breaking 
of R-parity ensures the breaking of lepton 
and/or baryon number. On the other hand,
allowing for breaking of R-parity opens up the possibility of 
generating Majorana mass terms for the neutrinos 
\cite{Roy-mu,Roy:1996bua,susybneu,
susybneu1,oneloop,twoloop,numssm,extrarightnu,gauginoseesaw,spontaneous}.
This can be done through one loop \cite{oneloop,susybneu1} 
and two loop \cite{twoloop} diagrams generated via the 
lepton number breaking trilinear couplings $\lambda$ and 
$\lambda'$ (see Eq. (\ref{eq:rpv})).  
Small neutrino masses can also be generated by the R-parity violating 
bilinear coupling $\hat H_u\hat L$ \cite{Roy-mu,Roy:1996bua} 
where $\hat H_u$ is the 
Higgs superfield and $\hat L$ is the leptonic superfield. 
This term  gives rise to 
higgsino-neutrino mixing, whereas    
the gaugino-neutrino mixing  will   be generated 
from the kinetic terms (which we give in Eq. (\ref{eq:lkinexp})),  
once the sneutrino fields get vacuum expectation values.
Hence this   $ \hat H_u\hat L $ bilinear coupling alongwith the 
gaugino-neutrino mixing terms bring about a neutralino-neutrino 
mixing and 
one can easily obtain a seesaw like formula for the light neutrinos, 
where the neutrino mass is suppressed by 
the neutralino mass. However, unlike the higher loop neutrino mass 
models with trilinear R-parity violating couplings, 
the bilinear R-parity breaking gaugino seesaw model can only 
generate one massive neutrino. Neutrino oscillation experiments have 
confirmed that we have at least two massive neutrinos. 
Therefore, one needs something extra in the gaugino seesaw models for 
generating the correct neutrino mass matrix which is viable 
with experiments.   
Models with explicit R-parity violation have been put forth where one 
introduces additional chiral superfields containing heavy 
right-handed neutrinos 
\cite{extrarightnu,numssm}, in addition to usual MSSM field content. 
However, in all models with R-parity violation, proton decay 
remains a severe constraint, and 
one has to explain why the 
lepton number violating terms in the R-parity violating part of the 
superpotential 
should be so much more larger 
compared to the one which involves only quarks 
(the $\lambda^{''}$ coupling in Eq. (\ref{eq:rpv})) 
and which should be 
heavily suppressed in order to suppress proton decay\footnote{Proton decay 
obviously requires simultaneous presence of baryon and lepton number 
violation. Lepton number is necessarily violated to generate Majorana 
neutrino masses. However, proton decay can be forbidden by forbidding 
the  $\lambda^{''}$ coupling.}. One would generally have to impose 
additional symmetry arguments in order to explain this.  

A possible way of circumventing this problem is by invoking  
spontaneous breaking of R-parity 
\cite{Rparityold2,gauginoseesaw,spontaneous,rmajoron,gauge1,
gauge2,pavel}. 
In this scenario, R-parity is 
conserved initially in the superpotential, and only once 
the sneutrino fields acquire vacuum expectation value, R-parity 
breaking terms are generated spontaneously. In presence of 
additional singlet or triplet matter chiral fields, this provides 
a natural explanation for the origin of the  
R-parity and lepton number violating 
bilinear term\footnote{It is possible to 
realize  the other terms $\lambda$ and
$\lambda^{\prime}$ from the R-parity 
conserving MSSM superpotential only
after redefinition of  basis\cite{Roy-mu}.} 
$\epsilon$, without generating 
the baryon number violating $\lambda^{\prime \prime}$ term 
in the superpotential. Therefore, one can generate neutrino 
masses without running into problems with proton decay in this class of 
supersymmetic models.  

There have been earlier attempts to construct spontaneous R-parity violating
models within the MSSM gauge group. 
In all these models the R-parity is broken spontaneously 
when the sneutrino acquires a VEV. This automatically breaks 
lepton number spontaneously. Since lepton number is a global 
symmetry, this might give rise to a massless Goldstone called the 
Majoron \cite{majoronoriginal,Rparityold2,
gauginoseesaw,spontaneous,rmajoron}. All models which 
predict presence of Majoron are severely constrained, unless 
the Majoron is mostly singlet. The phenomenology of these 
latter types of models has been studied in detail in 
\cite{spontaneous,rmajoron}. 
A possible way of avoiding the Majoron is by gauging the 
U(1) symmetry associated with lepton number such that the 
spontaneous R-parity and lepton number violation comes 
with the new gauge symmetry breaking. 
This gives rise to an additional neutral gauge boson and 
the phenomenology of these models have also been studied 
extensively in the literature \cite{gauge1,gauge2}. 
This idea has been used in a series of recent papers 
\cite{pavel} 


In this work we study  spontaneous R-parity violation in the presence of
a SU(2) triplet $Y=0$ matter chiral superfield, 
where we stick  to the gauge
group of the  
minimal supersymmetric standard model. Lepton number is 
broken explicitly in our model by the Majorana mass term of the 
heavy fermionic triplet, thereby circumventing the problem of the Majoron.
We break R-parity  spontaneously by giving vacuum expectation values 
to the 3 MSSM sneutrinos and the one 
additional sneutrino associated with the triplet which 
leads to the  lepton-higgsino and  lepton-gaugino mixing in addition to the
conventional Yukawa driven neutrino-triplet neutrino mixing.
This opens up the  possibility of  generating neutrino 
mass  from a combination of the conventional  Type-III seesaw and 
the gaugino seesaw. We restrict ourself 
to just one additional SU(2) triplet $Y=0$ matter chiral superfield,
and explore the possibility of getting viable neutrino mass splitting 
and the mixing angles at tree level. With one generation of heavy triplet 
we get two massive neutrinos while the third state remains 
massless. Like the neutralino sector 
we also  have R-parity conserving mixing between the 
standard model charged leptons and the heavy triplet 
charged lepton states. The spontaneous R-parity breaking  
brings about mixing between the charginos and the charged leptons, 
and hence modifies  the chargino mass spectrum. 
In addition to the usual charged leptons, our model contains a 
pair of heavy charged fermions coming from 
the fermionic component of the triplet superfield. 
Therefore, the charginos have contributions from these heavy 
exotic charged particles as well and we expect 
distinctive collider signatures due to this. 
Another novel feature of our  
model comes from the fact that the additional triplet fermions and 
sfermions have direct gauge interactions. Hence they offer a much richer
collider phenomenology. We  discuss in 
brief about the possibility of detecting our model at 
colliders and the predicted R-parity violating 
signatures. 

The main aspects of our spontaneous R-parity violating 
model compared to the ones already 
discussed in the literature are the following:
\begin{itemize}

\item We introduce one chiral superfield containing the triplets 
of SU(2) and with $Y=0$. The earlier 
viable models have considered one or more singlet 
chiral superfields containing the right-handed 
neutrino.

\item We have an explicit breaking of the lepton number 
due to the presence of the mass term of this chiral 
superfield in the superpotential. Therefore unlike 
as in \cite{Rparityold2,
gauginoseesaw,spontaneous,rmajoron},  
the spontaneous breaking of R-parity does not create any Majoron 
in our model. Since we do not have any additional gauge symmetry, 
we also do not have any additional neutral gauge boson 
as in \cite{gauge1,gauge2,pavel}. 

\item Since we have only one additional triplet chiral superfield 
we have two massive neutrinos, with the lightest one remaining 
massless. One of the neutrinos get mass due to type III seesaw and 
another due to gaugino seesaw. Combination of both gives 
rise to a neutrino mass matrix which is consistent with the 
current data. 

\item The triplet chiral superfield in our model modifies not 
only the neutrino-neutralino mass matrix, 
but also the charged lepton --  
chargino mass matrix. Being a triplet, it contains  
one neutral Majorana fermion $\Sigma^0$, 
two charged fermions $\Sigma^\pm$, one sneutrino $\tilde\Sigma^0$, 
and two charged sfermions $\tilde\Sigma^\pm$. 
Therefore, 
our neutral fermion mass matrix is a $8\times 8$ matrix, giving 
mixing between the gauginos, higgsinos as well as the new 
TeV-scale neutral fermion $\Sigma^0$. Likewise, the new 
charged fermions $\Sigma^\pm$ will mix with the charginos and 
the charged leptons.  

\item There are thus new TeV-mass neutral and charged leptons 
and charged scalars  
in our model, which will have mixing with other MSSM particles. 
These could be probed at future 
colliders and could lead to a rich phenomenology. 
We give a very brief outline of the 
collider signatures in this work. We plan 
to make a detailed study of the collider aspects of our 
model in a future work.

\end{itemize}

The paper is organized as follows. In section 2 we  describe the model and 
in section 3 we present the  symmetry breaking analysis. In section 4, we
discuss the 
neutralino-neutrino mass matrix and in section 5 we discuss the
chargino-charged lepton
mass matrix. We concentrate on the neutrino phenomenology in section 6 and
discuss the possibility
of getting correct mass splittings and mixings even with one generation of 
SU(2) triplet matter chiral supermultiplet. In section 7 we discuss about
detecting our model in 
colliders and finally in section 8 we present our conclusion. Discussion on
soft-supersymmetry breaking
terms, gaugino-lepton-slepton mixing and the exact analytic expression of
the low energy neutrino mass matrix have been
presented in Appendix A, Appendix B and in Appendix C, respectively. In
Appendix C we also analyze the 
constraints on the different sneutrino vacuum expectation values coming from
the neutrino mass scale.

\section{The Model}
The discrete $Z_2$ symmetry R-parity  is defined as $R_p=(-1)^{3(B-L)+2S}$
where $B$, $L$ represents the baryon number and lepton number of the particles
and $S$ is the spin. The matter chiral superfields have R-parity $-1$ whereas
the Higgs chiral superfields have R-parity +1. The R-parity conserving
superpotential of the MSSM is 
\be 
W_{MSSM}=Y_e\hat{H_d}\hat{L}\hat{E^c}+
Y_d\hat{H_d}\hat{Q}\hat{D^c}-Y_u\hat{H_u}\hat{Q}\hat{U^c}
+\mu \hat{H_u}\hat{ H_d}.
\label{eq:wmssm}
\ee
The other   terms which are allowed by supersymmetry invariance and gauge
invariance but violate R-parity are 
\be
W_{\not{R_p}}= -\epsilon_i \hat{H_u} \hat{L_i}+ \lambda_{ijk}
\hat{L_i}\hat{L_j}\hat{E^c_k}+
{\lambda_{ijk}^{\prime}}\hat{L_i}\hat{Q_j}\hat{D^c_k}+{\lambda_{ijk}^{\prime
\prime}}\hat{U^c_i}\hat{D^c_j}\hat{D^c_k}.
\label{eq:rpv}
\ee
As already mentioned in the introduction, in this work we will explore
spontaneous R-parity violation in the presence of SU(2) triplet $Y=0$  matter
chiral superfield.
The matter chiral supermultiplets of the model   are:\\
\\
\vspace*{0.3cm}
$\hat{Q}=\pmatrix{\hat{U} \cr \hat{D}}$, $\hat{L}=\pmatrix{\hat{\nu} \cr \hat{E}}$,
$\hat{U^c}$, $\hat{D^c}$ and $\hat{E^c}$\\
and the Higgs chiral supermultiplets are :\\
\\
\vspace*{0.4cm}
$\hat{H_u}=\pmatrix{\hat{H}_u^+ \cr \hat{H}_u^0}$, $\hat{H_d}=\pmatrix{\hat{H}_d^0 \cr \hat{H}_d^{-}}$.\\
In addition to the standard supermultiplet contents of the MSSM we introduce  one  SU(2) triplet matter chiral supermultiplet $\hat{{\Sigma}}_R^c$ with $U(1)_Y$ hypercharge $Y=0$. We represent $\hat{{\Sigma}}_R^c$ as 
\be
\hat{{\Sigma}}_R^c=\frac{1}{\sqrt{2}}\pmatrix{{{\hat{\Sigma}_R}^{0c}}  & \sqrt{2}\hat{{\Sigma}}_R^{-c} \cr
\sqrt{2}\hat{{\Sigma}}_R^{+c} & -{{\hat{\Sigma}_R}^{0c}}  }.
\label{eq:sigmul}
\ee
The three different chiral superfields in this multiplet are 
\be 
\hat{{\Sigma}}_R^{+c} ={\tilde{{\Sigma}}_R^{+c}}+\sqrt{2}\theta {\Sigma_R^{+c}}+ \theta \theta F_{{\Sigma_R^{+c}}},
\ee
\be
\hat{{\Sigma}}_R^{-c} ={\tilde{{\Sigma}}_R^{-c}}+\sqrt{2}\theta {\Sigma_R^{-c}}+ \theta \theta F_{{\Sigma_R^{-c}}},
\ee
\be
{\hat{\Sigma}}_R^{0c}= \tilde{{\Sigma}}_R^{0c}+\sqrt{2}\theta {\Sigma_R^{0c}}+ \theta \theta F_{{\Sigma_R^{0c}}}.
\ee
The SU(2) triplet fermions are ${\Sigma_R^{+c}}$, ${\Sigma_R^{-c}}$ and ${\Sigma_R^{0c}}$ and their scalar superpartners are represented as ${\tilde{{\Sigma}}_R^{+c}}$, ${\tilde{{\Sigma}}_R^{-c}}$ and $\tilde{{\Sigma}}_R^{0c}$ respectively. $F_{{\Sigma_R^{+c}},{\Sigma_R^{-c}},{\Sigma_R^{0c}}}$ represent the different auxiliary fields of the above mentioned multiplet. R-parity of $\hat\Sigma_R^c$ is $-1$ where componentwise the fermions ${\Sigma_R^{+c}}$, ${\Sigma_R^{-c}}$ and ${\Sigma_R^{0c}}$  have $R_p=+1$ and their scalar superpartners have $R_p=-1$. With these field contents, the R-parity conserving superpotential $W$ will be 
\be 
W=W_{MSSM}+W_{\Sigma},
\ee
where $W_{MSSM}$ has already been written  in  Eq. (\ref{eq:wmssm}) and $W_{\Sigma}$ is given by
\be
W_{\Sigma}= -{Y_{\Sigma_i}}\hat{H_u}^T (i\sigma_2) \hat{{\Sigma}}_R^c{\hat{L}}_i+ \frac{M}{2} Tr[\hat{{\Sigma}}_R^c\hat{{\Sigma}}_R^c].
\label{eq:wwsigma}
\ee
$W_{\Sigma}$ is  clearly R-parity conserving. The scalar fields $\tilde{{\Sigma}}_R^{0c}$ and ${\tilde{\nu}_{L_i}}$ are odd under R-parity. Hence in this model, R-parity would be spontaneously broken by the vacuum expectation values of these sneutrino fields. We will analyze the potential and spontaneous R-parity violation in the next section. On writing explicitly, one will get these following few terms from the above superpotential $W_{\Sigma}$,
\be
W_{\Sigma}&=& \frac{{Y_{\Sigma_i}}}{\sqrt{2}} \hat{H}_u^0 \hat{\Sigma}_R^{0c} \hat{\nu}_{L_i} +{Y_{\Sigma_i}} \hat{H}_u^0\hat{\Sigma}_R^{-c} \hat{l_i}^-+\frac{{Y_{\Sigma_i}}}{\sqrt{2}} \hat{H}_u^+\hat{\Sigma}_R^{0c}\hat{l_i}^- -{Y_{\Sigma_i}} \hat{H}_u^+ \hat{\Sigma}_R^{+c}\hat{\nu}_{L_i} \nn \\
& & +\frac{M}{2} \hat{\Sigma}_R^{0c}\hat{\Sigma}_R^{0c} + M\hat{\Sigma}_R^{+c}\hat{\Sigma}_R^{-c}. 
\label{eq:wsigmaexp}
\ee
The kinetic terms of the $\hat{\Sigma}_R^c $ field is 
\be 
L_{\Sigma}^{k}= \int d^4\theta  Tr[\hat{\Sigma _R^c}^{\dagger}e^{2gV} \hat{\Sigma_R^c}].
\label{eq:ssigmak}
\ee
The soft supersymmetry  breaking Lagrangian of this model is  
\be
L^{\rm{soft}}=L_{\rm{MSSM}}^{\rm{soft}}+L_{\Sigma}^{\rm{soft}}.
\ee
For completeness  we write the $L_{\rm{MSSM}}^{\rm{soft}}$ Lagrangian in the
Appendix A. $L_{\Sigma}^{\rm{soft}}$ contains the supersymmetry breaking terms corresponding to scalar $\tilde{{\Sigma}}_R^c$ fields
and is given by 
\be
L_{\Sigma}^{\rm{soft}}=
-m_{\Sigma}^2 \rm{Tr}[ \tilde{\Sigma}_R^{c^{\dagger}}
\tilde{{{\Sigma_R^c}}}]-({\tilde{m}}^2\rm{Tr}[\tilde{{{\Sigma_R^c}}}\tilde{{{\Sigma_R^c}}}]+h.c)-({A_{\Sigma_i}}{\it
  H_u^T}i\sigma_2\tilde\Sigma_R^c \tilde{L}_i+h.c),
\label{eq:lsigmasoft}
\ee
where 
\be
\tilde{\Sigma}_R^c =\frac{1}{\sqrt{2}}\pmatrix{ {\tilde{\Sigma}_R^{0c}}  & \sqrt{2} \tilde{\Sigma}_R^{-c} \cr 
\sqrt{2}\tilde{\Sigma}_R^{+c} & -\tilde{\Sigma}_R^{0c}}.\ee
We explicitly show  in the Appendix A all the possible trilinear terms which
will be generated from Eq. (\ref{eq:wsigmaexp}) and the interaction terms
between gauginos and SU(2) triplet fermion and sfermion coming from Eq. (\ref{eq:ssigmak}).  In the next section we analyze the neutral component of the potential and discuss  spontaneous R-parity violation through $\tilde{\Sigma}_R^{0c}$ and $\tilde{\nu}_{L_i} $ vacuum expectation values.

\section{Symmetry Breaking}

In this section we write down the scalar potential which will be relevant to analyze the symmetry breaking of the Lagrangian. The potential  is  
\be
V=V_F+V_D+V_{soft},
\ee
where $ V_F$ and $V_D$, the contributions from different auxiliary components of the  chiral superfield and different vector supermultiplets  are  given by 
\be 
V_F=  \sum_kF_k^*F_k \\
V_D=\frac{1}{2}\sum_a D^aD^a
\ee respectively. Here the index $k$  denotes all possible auxiliary components of the matter chiral superfields whereas the index $a$ is the gauge index. The contribution from the soft supersymmetry breaking Lagrangian is given by $V_{soft}$. Below we write down the neutral component of the potential which would be relevant for our symmetry breaking analysis. The neutral component of the potential is given by
\be 
V_{neutral}=V_F^n+ V_D^n+ V_{soft}^n  ,
\ee
where 
\be 
V_F^n=|F_{H_u^0}|^2+|F_{H_d^0}|^2+|F_{ {\tilde{\nu}}_{L_i}}|^2+ |F_{{\tilde{\Sigma}_R^{0c}}}|^2.
\ee
The different $F_k$ are given by 
\be
F_{H_u^0}^*= \mu H_d^0 - \sum_i\frac{{Y_{\Sigma_i}}}{\sqrt{2}} \tilde{\Sigma}_R^{0c}\tilde{\nu}_{L_i}+...,
\ee

\be
F_{H_d^0}^*= \mu H_u^0+...,
\ee
\be
{F}^*_{\tilde{\Sigma}_R^{0c}}= - \sum_i\frac{{Y_{{\Sigma_i}}}}{\sqrt{2}} H_u^0\tilde{\nu}_{L_i} -M\tilde{\Sigma}_R^{0c}+...,
\ee
\be
F^*_{\tilde{\nu}_{L_i}}= - \frac{{Y_{{\Sigma_i}}}}{\sqrt{2}} H_u^0\tilde{\Sigma}_R^{0c}+... .
\ee
In the above equations $...$ represents other terms which will not contribute to the neutral component of the potential.  With all these $F_k$'s, $V^n_F$ is given by 
\be 
V_F^n= |\mu H_d^0|^2+ |\mu H_u^0|^2+\frac{1}{2}|\sum_i{{{Y_{\Sigma_i}}}\tilde{{\Sigma}}_R^{0c}\tilde{{\nu}}_{L_i}}|^2+\frac{1}{2}\sum_i{|{{Y_{\Sigma_i}}}  H_u^0\tilde{{\Sigma}}_R^{0c}|^2}+\frac{1}{2}|\sum_i{{{Y_{\Sigma_i}}} H_u^0\tilde{{\nu}}_{L_i}}|^2 \\ - [ \mu H_d^0 (\sum_i{ \frac{{Y_{\Sigma_i}}}{\sqrt{2}} \tilde{{\Sigma}}_R^{0c} \tilde{{\nu}}_{L_i}})^*+c.c] + |M|^2 \tilde{\Sigma}_R^{{0c}^{*}} \tilde{{\Sigma}}_R^{0c}+[\sum_i{ \frac{{Y_{\Sigma_i}}}{\sqrt{2}} H_u^0 \tilde{{\nu}}_{L_i}}(M \tilde{{\Sigma}}_R^{0c})^*+c.c] .
\ee
As we have three generation of leptons hence the generation index $i$ runs as  {$i$=1,2,3}.  The D term contribution of $V_{neutral}$ is  given as 
\be
V_D^n=\frac{1}{8}(g^2+{g^{\prime}}^2)(|H_d^0|^2-|H_u^0|^2+\sum_i{|\tilde{\nu}_{L_i}|^2})^2.
\label{eq:vd}
\ee
Note that $\tilde{{\Sigma}}_R^{0c}$ which  has $Y=0$ and the third component of the isospin $T_3=0$ does not  contributes to $V_D^n$. 
The soft supersymmetry breaking contributions to the neutral part of the potential is given by $V_{soft}^n$ where 
\be
V_{soft}^n &=& -(bH_u^0H_d^0+c.c)+m_{H_u}^2 |H_u^0|^2+m_{H_d}^2 |H_d^0|^2 \\ \nn
 && +m_\Sigma^2 \tilde{\Sigma}_R^{0c^{*}}\tilde{\Sigma}_R^{0c}+ [{\tilde{m}}^2\tilde{\Sigma}_R^{0c}\tilde{\Sigma}_R^{0c}+c.c]\\  \nn
 && + \sum_{i}{m^2_{\tilde{\nu_i}}} {\tilde{\nu}_{L_i}}^{*}\tilde{\nu}_{L_i}+[\sum_i \frac{A_{\Sigma_i}}{\sqrt{2}}H_u^0 \tilde{\Sigma}_R^{0c}\tilde{\nu}_{L_i}+c.c].
\label{eq:vsoft}
\ee
We represent  the vacuum expectation values of  $H_u^0$, $H_d^0$, $\tilde{\nu}_{L_i}$ and $ \tilde{\Sigma}_R^{0c}$ as  $\langle H_u^0 \rangle=v_2$, $\langle H_d^0 \rangle=v_1$, $\langle  \tilde{\nu}_{L_i}  \rangle=u_i$ and $\langle \tilde{\Sigma}_R^{0c}\rangle =\tilde{u}$. We have considered a diagonal $ m^2_{\tilde{\nu}} $ matrix. In terms of these vacuum expectation values the neutral component of the potential is 
\be
\langle V_{neutral}\rangle &=& (|\mu|^2+m_{H_u}^2)|v_2|^2+(|\mu|^2+m_{H_d}^2)|v_1|^2-(bv_1v_2+c.c) \nn \\
&& +\frac{1}{8}(g^2+{g^{\prime}}^2){(|v_1|^2-|v_2|^2+\sum_i{|u_i|^2})}^2  +(|M|^2+m_{\Sigma}^2)|\tilde{u}|^2 \nn \\
&& + \sum_im_{\tilde{\nu_i}}^2|u_i|^2+ [{\tilde{m}}^2\tilde{u}^2+c.c]+\frac{1}{2}|\sum_i{Y_{\Sigma_i}}\tilde{u}u_i|^2+ \frac{1}{2}\sum_i{|{Y_{\Sigma_i}}|^2}|\tilde{u}v_2|^2 \nn \\ 
&& +\frac{1}{2} |\sum_i Y_{\Sigma_i}u_iv_2|^2 +(\sum_i\frac{A_{\Sigma_i}}{\sqrt{2}}v_2u_i\tilde{u}+c.c)-(\mu v_1(\sum_i \frac{Y_{\Sigma_i}}{\sqrt{2}}\tilde{u}u_i)^*+c.c)\nn \\
&& +[\sum_i \frac{Y_{\Sigma_i}}{\sqrt{2}}v_2u_i(M\tilde{u})^*+c.c].
\label{eq:vneub}
\ee
For simplicity we assume all the vacuum expectation values and all the parameters are real. Hence $\langle V_{neutral}\rangle$ simplifies to 
\be
\langle V_{neutral}\rangle &=& (\mu^2+m_{H_u}^2)v_2^2+(\mu^2+m_{H_d}^2)v_1^2-2bv_1v_2+\frac{1}{8}(g^2+{g^{\prime}}^2)(v_1^2-v_2^2+\sum_i{u_i^2})^2 \nn  \\
&&  +(M^2+m_{\Sigma}^2)\tilde{u}^2+ \sum_im_{\tilde{{\nu_i}}}^2u_i^2+ 2{\tilde{m}}^2\tilde{u}^2+\frac{1}{2}(\sum_i{Y_{\Sigma_i}}u_i)^2\tilde{u}^2+\frac{1}{2}\sum_i {(Y_{\Sigma_i})^2}\tilde{u}^2v_2^2 \nn \\ 
&&+ \frac{1}{2}(\sum_i{Y_{\Sigma_i}}u_i)^2v_2^2 +\sqrt{2}\sum_i({A_{\Sigma_i}}v_2u_i\tilde{u}- \mu v_1{Y_{\Sigma_i}}\tilde{u}u_i+{Y_{\Sigma_i}}M v_2u_i\tilde{u}).
\label{eq:vneua}
\ee
Minimizing $\langle V_{neutral} \rangle$ with respect to  $v_1$, $v_2$, $\tilde{u}$ and $u_i$  we get the following four equations, 
\be
2(\mu^2+m_{H_d}^2)v_1-2bv_2+\frac{v_1}{2}(g^2+{g^{\prime}}^2)(v_1^2-v_2^2+\Sigma_i{u_i^2})-\sqrt{2}\mu \tilde{u} \sum_i{{Y_{\Sigma_i}}u_i}=0,
\label{eq:mina}
\ee
\be
2(\mu^2+m_{H_u}^2)v_2-2bv_1-\frac{v_2}{2}(g^2+{g^{\prime}}^2)(v_1^2-v_2^2+\sum_i{u_i}^2)+v_2(({\sum_i{Y_{\Sigma_i}}u_i})^2
+\sum_i{{{(Y_{\Sigma_i}})}^2}\tilde{u}^2) \nn \\ 
+\sqrt{2}\sum_i{({A_{\Sigma_i}}+{Y_{\Sigma_i}}M)u_i} \tilde{u}=0,
\label{eq:minb}
\ee
\be
2(m_{{\Sigma}}^2+2M^2+2{\tilde{m}}^2)\tilde{u}+(\sum_i{{Y_{\Sigma_i}}u_i})^2\tilde{u}+\sqrt{2}\sum_i{({A_{\Sigma_i}}v_2u_i-\mu
  {Y_{\Sigma_i}}v_1u_i + {Y_{\Sigma_i}}Mv_2u_i)} \nn \\ 
 +\sum_i{{{(Y_{\Sigma_i}})}^2}\tilde{u}v_2^2  =0 ,
\label{eq:mind}
\ee
\be
\frac{u_i}{2}(g^2+{g^{\prime}}^2)(v_1^2-v_2^2+\sum_j{u_j^2})+(v_2^2+\tilde{u}^2)[{{Y_{\Sigma_i}}}^2u_i+{Y_{\Sigma_i}}\sum_{j
  \neq i}{{Y_{\Sigma_j}}u_j}] + \sqrt{2}{A_{\Sigma_i}}v_2\tilde{u} \nn \\ 
+\sqrt{2}[{Y_{\Sigma_i}}Mv_2\tilde{u}-\mu {{Y_{\Sigma_i}}}v_1\tilde{u}]+2m_{\tilde{\nu_i}}^2u_i=0.
\label{eq:minc}
\ee
respectively. In the last equation  note that the index $i$ is {\it not summed
  over}. As mentioned before, since $\tilde{\nu}_{L_i}$ and $\tilde{\Sigma}_R^{0c}$ have nontrivial R-parity, hence R-parity is spontaneously broken when $\tilde{\nu}_{L_i}$ and $\tilde{\Sigma}_R^{0c}$ take vacuum expectation values. As a result of this spontaneous R-parity violation, the  bilinear term $LH_u$ which will contribute to the neutrino mass matrix is generated. We will discuss in detail  about the neutralino-neutrino and chargino-charged lepton mass matrix in the next section. 

The minimization conditions given in Eqs. (\ref{eq:mina})-(\ref{eq:minc})
can be used to give constraints on the vacuum expectation values 
$u_i$ and $\tilde u$. In order to get such relations we 
drop the generation indices for the moment and consider $u_i=u$ 
for simplicity. 
In this case ${{Y_{\Sigma}}}$, ${{A_{\Sigma}}}$ and
$m_{\tilde{\nu_i}}^2$ contain no generation index and would be just
three numbers. From the simplified version of the last two equations
Eq. (\ref{eq:mind}) and Eq. (\ref{eq:minc}) it can then be shown that 
in the limit that $u$ is small\footnote{We will show later that 
small $u$ is demanded from the smallness of neutrino mass.} the two
$R_p$ breaking vacuum expectation values 
$u$ and $\tilde{u}$ are proportional to each other. 
Combining these two equations one gets
\be
\tilde{u}^2=\frac{u^2}{{Y_{\Sigma}}^2v_2^2+
2(m_{\Sigma}^2+2{\tilde{m}}^2+2M^2)}[\frac{1}{2}(g^2+g'^2)
(v_1^2-v_2^2+u^2)+2m_{\tilde{\nu}}^2+{Y_{\Sigma}}^2v_2^2].
\label{eq:uut}
\ee
Hence for small $u$ which is demanded from the smallness 
of neutrino mass (see the next section and Appendix C), 
$\tilde{u}$ will also be of the same order as $u$
unless there is a cancellation between the terms in the
denominator. In this work we will stick to the possibility of small $u$ and
$\tilde{u}$. We will show in the next section that one needs $u \sim
10^{-3}$ GeV to explain the neutrino data. Hence 
$\tilde{u}$ will also have to be $10^{-3}$ GeV.
In the  $u=0$ limit, $\tilde{u}$ would also be 0 and
this is our usual R-parity conserving  scenario.


\section{Neutralino-Neutrino Mass Matrix}

In this section we discuss the consequence of R-parity violation through the
neutralino-neutrino mixing. It is well known \cite{neuneu} that R-parity
violation results in mixing between the neutrino-neutralino states. In our
model the neutrino sector is enlarged and includes both the standard model
neutrino $\nu_{L_i}$,  as well as the heavier neutrino state
${\Sigma}_R^{0c}$,  which is a component of SU(2) triplet superfield. Since
R-parity is violated we get  higgsino-neutrino mixing terms
$\frac{Y_{\Sigma_i}}{\sqrt{2}}\tilde{u} \tilde{H}_u^0{{\nu_{L_i}}}$ and
$\frac{Y_{\Sigma_i}}{\sqrt{2}}u \tilde{H}_u^0 {\Sigma}_R^{0c}$, in addition to
the conventional R-parity conserving Dirac mass term
$\frac{Y_{\Sigma_i}}{\sqrt{2}}v_u {{\Sigma}}_R^{0c}{{\nu_{L_i}}}$.  The
R-parity breaking former two terms  originated  from  the term
$\frac{Y_{\Sigma_i}}{\sqrt{2}}
\hat{H_u^0}\hat{{\Sigma}}_R^{0c}{\hat{\nu}_{L_i}}$ in
Eq. (\ref{eq:wsigmaexp}), once the sneutrino fields $\tilde{\nu}_{L_i}$ and
$\tilde{\Sigma}_R^{0c}$ get vacuum expectation values. The third term also has
the same origin  and it is the conventional Dirac mass term in Type-I or
Type-III seesaw. In addition to the higgsino-neutrino mixing terms generated
from the superpotential $W_{\Sigma}$, there would also be gaugino-neutrino
mixing terms generated from the kinetic term of the $\hat{L_i}$ and
$\hat{\Sigma}_R^c$. In the Appendix B we show explicitly the contributions
coming from $W_{\Sigma}$ and the neutrino-sneutrino-gaugino terms originating
from the kinetic term of the triplet superfield $\hat{\Sigma}_R^c$ written
down in Eq. (\ref{eq:ssigmak}). Here  we write the color singlet
neutral-fermion mass matrix of this model in the basis
$\psi=(\tilde{\lambda}^0\rm{,}\tilde{\lambda}^3 \rm{,}\tilde{H}_d^0 \rm{,}\tilde{H}_u^0\rm{,}
\Sigma_R^{0c} \rm{,} \nu_{L_1} \rm{,} \nu_{L_2}\rm{,} \nu_{L_3})^T$ where with
one generation of ${\Sigma_R^c}$,  the neutral fermion mass matrix is a   $8
\times 8$  matrix. The mass term is given by 
\be
L_n=-\frac{1}{2} \psi^T M_n \psi+h.c.
\ee
where 
\be
M_n=\frac{1}{\sqrt{2}}\pmatrix{ \sqrt{2}M^1 & 0 & -{g^{\prime}v_1}& {g^{\prime}v_2}& 0& -{g^{\prime}u_1}&-{g^{\prime}u_2}&-{g^{\prime}u_3} \cr
0& \sqrt{2} M^2 & {g v_1}& -{g v_2}& 0 & {gu_1}& {gu_2}& {gu_3}\cr
-{g^{\prime}v_1}&{g v_1}& 0 & -\sqrt{2}\mu & 0 & 0 & 0 & 0 \cr
{g^{\prime}v_2} & -{g v_2} & -\sqrt{2}\mu & 0 & \sum_{i}{{Y_{\Sigma_i}}u_i}& {Y_{\Sigma_1}}\tilde{u}&
{Y_{\Sigma_2}}\tilde{u}&{Y_{\Sigma_3}}\tilde{u}\cr
0 & 0 & 0 & \sum_{i}{{Y_{\Sigma_i}}u_i} & \sqrt{2}M & {Y_{\Sigma_1}}v_2& {Y_{\Sigma_2}}v_2& {Y_{\Sigma_3}}v_2\cr
-{g^{\prime}u_1}& {g u_1}&0 & {Y_{\Sigma_1}}\tilde{u} &{Y_{\Sigma_1}}v_2&  0 & 0 & 0\cr
-{g^{\prime}u_2}& {g u_2}&0 & {Y_{\Sigma_2}}\tilde{u} &{Y_{\Sigma_2}}v_2&  0 & 0 & 0\cr
-{g^{\prime}u_3}&{g u_3}&0 & {Y_{\Sigma_3}}\tilde{u} &{Y_{\Sigma_3}}v_2&  0 & 0 & 0\cr
}.
\label{eq:massnm}
\ee
Here $M^{1,2}$ are the soft supersymmetry breaking gaugino mass parameters
(see Appendix B), whereas $M$ corresponds to the triplet-fermion bilinear term. 
We define the $3\times 5 $ matrix $m_D$ as
\be
m_D^T=\frac{1}{\sqrt{2}}\pmatrix{ -{g^{\prime}u_1}& {g u_1}&0 & {Y_{\Sigma_1}}\tilde{u} &{Y_{\Sigma_1}}v_2 \cr
-{g^{\prime}u_2}& {g u_2}&0 & {Y_{\Sigma_2}}\tilde{u} &{Y_{\Sigma_2}}v_2 \cr
-{g^{\prime}u_3}& {g u_3}&0 & {Y_{\Sigma_3}}\tilde{u} &{Y_{\Sigma_3}}v_2
}.
\label{eq:mdt}
\ee
Defined in this way, the  $8 \times 8$ neutral fermion mass matrix can be written as 
\be 
M_n=\pmatrix{M^{\prime} & m_D \cr m_D^T & 0 },
\ee
where $M^{\prime}$ represents the $5 \times 5 $ matrix
\be
M^{\prime}= \frac{1}{\sqrt{2}}\pmatrix{ \sqrt{2}M^1 & 0 & -{g^{\prime}v_1}& {g^{\prime}v_2}& 0 \cr
0& \sqrt{2}M^2 & {g v_1}& -{g v_2}& 0 \cr
-{g^{\prime}v_1}& {g v_1}& 0 & -\sqrt{2}\mu & 0  \cr
{g^{\prime}v_2} & -{g v_2} & -\sqrt{2}\mu & 0 & \sum_{i}{{Y_{\Sigma_i}}u_i} \cr
0 & 0 & 0 & \sum_{i}{{Y_{\Sigma_i}}u_i} & \sqrt{2}M }.
\label{eq:masnm}
\ee
The low energy neutrino mass would be generated once the neutralino and exotic triplet fermions get  integrated out. Hence the low energy neutrino mass matrix  $M_{\nu}$ is
\be
M_{\nu} \sim  -m_D^T {M^{\prime}}^{-1}m_D.
\ee
For $M^{\prime} $ in the TeV range,  $M_{\nu} \sim $ 1 eV demands that $m_D$
should be  $10^{-3}$ GeV.  If one takes $v_2 \sim 100$ GeV  then this sets
$Y_{\Sigma} \sim 10^{-5}$ and  the scale of $u$ to be $10^{-3}$ GeV. Since in
our model for small value of $u$, the $\tilde{u}$ and $u$ are proportional to
each other, hence we naturally get $\tilde{u} \sim u \sim 10^{-3}$ GeV. We have discussed in more detail in Appendix C how the smallness of neutrino mass can restrict the  vacuum expectation values  $u_i$, $\tilde{u}$ and the Yukawas ${Y_{\Sigma_i}}$.  One can clearly see from Eq. (\ref{eq:massnm}) that in the $u=0$ and $\tilde{u}=0$ limit, the gaugino-higgsino sector completely decouples from the standard model neutrino-exotic neutrino sector and the low energy neutrino  mass would be governed via the usual Type-III seesaw only.

\section{Chargino-Charged Lepton Mass Matrix}

Like the neutralino-neutrino mixing as discussed in the previous section, R-parity violation will also result in chargino-charged lepton mixing, which in our model is significantly different compared to the other existing models of spontaneous and explicit R-parity violation, because of the presence of extra heavy triplet charged fermionic states in our model. Like the enlarged neutrino sector ($\nu_{L_i}$,$\Sigma_R^{0c}$) we have also an extended charged lepton sector. With one generation of $\hat{\Sigma}_R^c$ we have two additional heavier triplet charged leptons $\Sigma_R^{+c}$ and $\Sigma_R^{-c}$ in our model, in addition to the standard model charged leptons.  Hence we get mixing between the charginos and the standard model charged leptons as well as the heavier triplet charged leptons.  The possible contributions to the different mixing terms would come from  the superpotential as well as from the kinetic terms of the different superfields. Since we have written down explicitly the charginos-charged lepton-sneutrino interaction terms in Appendix, it is straightforward to see the contribution to the mass matrix coming from Eq. (\ref{eq:fsfg2}), Eq. (\ref{eq:fsfg3}) and Eq. (\ref{eq:lkinexp})  once the $\tilde{\Sigma}_R^{0c}$ and $\tilde{\nu}_{L_i}$ states get vacuum expectation values.  The chargino-charged lepton mass matrix in the basis  $\psi_1= ({ \tilde{\lambda}^+ \rm{,} \tilde{H}_u^+  \rm{,} l_{1}^c  \rm{,} l_{2}^c  \rm{,} l_{3}^c\rm{,} \Sigma_R^{-c}})^T$ and 
 $\psi_2= ({ \tilde{\lambda}^-  \rm{,} \tilde{H}_d^- \rm{,} l_1\rm{,} l_{2}\rm{,} l_{3}  \rm{,} \Sigma_R^{+c}})^T$  is 

\be
L_{c}=-\psi_1^TM_c \psi_2 +h.c,
\ee
where

\be
M_c=\frac{1}{\sqrt{2}}\pmatrix{ \sqrt{2}M^2 &  \sqrt{2}gv_1 &  \sqrt{2}gu_1 & \sqrt{2}gu_2 & \sqrt{2}gu_3    & {g\tilde{u}} \cr
 \sqrt{2}gv_2 &  \sqrt{2}\mu & {Y_{\Sigma_1}}\tilde{u} &  {Y_{\Sigma_2}}\tilde{u}& {Y_{\Sigma_3}}\tilde{u}&  -\sum_{i}{{\sqrt{2}Y_{\Sigma}}_iu_i} \cr
0 & - \sqrt{2}{Y_{e_1}}u_1 &  \sqrt{2}{Y_{e_1}}v_1 & 0 & 0 & 0\cr
0 & - \sqrt{2}{Y_{e_2}}u_2 & 0 &   \sqrt{2}{Y_{e_2}}v_1  & 0 & 0\cr
0 & - \sqrt{2}{Y_{e_3}}u_3 & 0 & 0 &  \sqrt{2}{Y_{e_3}}v_1 &  0 \cr
 -{g\tilde{u}} & 0 & \sqrt{2}{{Y_{\Sigma_1}}v_2} & \sqrt{2}{{Y_{\Sigma_2}}v_2} & \sqrt{2}{{Y_{\Sigma_3}}v_2} &  \sqrt{2}M
}.
\label{eq:chmssm}
\ee

The chargino-charged lepton mass matrix would be diagonalized by bi-unitary transformation $M_c=TM_c^dS^{\dagger}$.

\section{Neutrino Mass and Mixing}

R-parity violation can contribute significantly to the $3\times 3$  standard
model light neutrino mass matrix. In this section we concentrate on 
determining the neutrino mass square differences and the appropriate
mixings. It is well known that with only one generation of singlet/triplet
heavy Majorana neutrino it is not possible to get viable neutrino mass square
differences and mixings  in the 
R-parity conserving  Type-I or Type-III seesaw scenario. Since R-parity is
violated, we have neutrino-neutralino mixing apart from the conventional
standard model neutrino-heavy neutrino mixing, which has significant effect in
determining the low energy neutrino mass square differences and mixing angles
of PMNS mixing matrix\footnote{The standard charged lepton mass matrix 
which is obtained from Eq. (\ref{eq:chmssm}) turns out to be almost 
diagonal and therefore the PMNS mixing comes almost 
entirely from $M_\nu$.}, 
through the gaugino and higgsino  mass parameters
$M^{1,2}$, $\mu$ and the R-parity violating sneutrino vacuum expectation
values $u_i$ and $\tilde{u}$.
Below we write the approximate  $3 \times 3 $ standard model neutrino mass
matrix. Since ${Y_{\Sigma}}_i$, $u_i$ and $\tilde{u}$ are very small,  all the
terms which are proportional to $Y_{\Sigma_i}^2u_i^2$ and the terms
$Y_{\Sigma_i}^3 u_i \tilde{u}$ are neglected and we write down only the leading
order terms. The exact analytical expression of the low energy neutrino mass
matrix for our model has been given in the Appendix C. The approximate light
neutrino mass matrix has the following form,
\be 
M_{\nu} \sim \frac{v_2^2}{2M}A  + \frac{\alpha \mu}{2}B 
+ \frac{\alpha \tilde{u}v_1}{2\sqrt{2}}C ,
\label{eq:mnuapprox}
\ee
where the matrix $A$, $B$ and $C$ respectively are,
\be
 A =\pmatrix { Y_{\Sigma_1}^2 & Y_{\Sigma_1} Y_{\Sigma_2} 
& Y_{\Sigma_1} Y_{\Sigma_3} \cr
Y_{\Sigma_1} Y_{\Sigma_2} &  Y_{\Sigma_2}^2 & Y_{\Sigma_2} Y_{\Sigma_3} \cr
Y_{\Sigma_1} Y_{\Sigma_3} & Y_{\Sigma_2} Y_{\Sigma_3} & Y_{\Sigma_3}^2 },
\label{eq:A}
\ee
\be
B=\pmatrix { u_1^2 & u_1u_2 & u_1u_3 \cr
  u_1u_2 & u_2^2  & u_2u_3 \cr
  u_1u_3 & u_2u_3 & u_3^2 },
\label{eq:B}
\ee
\be
C =\pmatrix{2u_1Y_{\Sigma_1} & u_1Y_{\Sigma_2}+
u_2Y_{\Sigma_1}& u_1Y_{\Sigma_3}+u_3Y_{\Sigma_1}  \cr
u_1Y_{\Sigma_2}+u_2Y_{\Sigma_1} & 2u_2Y_{\Sigma_2} & u_2Y_{\Sigma_3}+u_3Y_{\Sigma_2} \cr
u_1Y_{\Sigma_3} +u_3Y_{\Sigma_1} & u_2Y_{\Sigma_3} +
u_3Y_{\Sigma_2} & 2u_3Y_{\Sigma_3} }.
\label{eq:C}
\ee
The parameter $\alpha$ depends on gaugino masses $M^{1,2}$, the higgsino mass
parameter $\mu$ and two vacuum expectation values $v_{1,2}$ as follows
\be
\alpha= \frac{(M_1g^2+M_2{g^{\prime}}^2)}{M_1M_2\mu-
(M_1g^2+M_2{g^{\prime}}^2)v_1v_2}.
\ee
Note that the 1st term in Eq. (\ref{eq:mnuapprox}) which depends only on the
Yukawa couplings ${Y_{\Sigma_i}}$, triplet fermion mass parameter $M$ and the
vacuum expectation value  $v_2$,  is the conventional R-parity conserving
Type-I or Type-III seesaw term. The 2nd and 3rd terms involve the gaugino mass
parameters $M^{1,2}$, the higgsino mass parameter $\mu$ and R-parity violating
vacuum expectation values  $u_i$ and $\tilde{u}$. Hence the appearance of
these two terms are undoubtedly the artifact of R-parity violation. 

We next discuss the neutrino oscillation parameters, the 
three angles in the
$U_{PMNS}$ mixing matrix and two mass square differences $\Delta m^2_{21}$ and
$\Delta m^2_{31}$. Note that with the mass matrix $M_{\nu}$ given in
Eq. (\ref{eq:A}), i.e taking only the effect  of  triplet 
Yukawa contribution into account  one would get the three following 
mass eigenvalues for the three
light standard model  neutrinos,
\be
  m_1=0,~~~ m_2=0,~~~  m_3=\frac{v_2^2}{2M}(Y_{\Sigma_1}^2+ 
Y_{\Sigma_2}^2+ Y_{\Sigma_3}^2). \nn
\label{eq:lowval}
\ee
 and the eigenvectors 
\be
\frac{1}{\sqrt{Y_{\Sigma_1}^2+ Y_{\Sigma_3}^2}}
\pmatrix{{-{Y_{\Sigma_3}}} \cr 0 \cr {{Y_{\Sigma_1}}}},~~~
\frac{1}{\sqrt{Y_{\Sigma_1}^2+ Y_{\Sigma_2}^2}}
\pmatrix{{-{Y_{\Sigma_2}}} \cr {{Y_{\Sigma_1}}} \cr 0},~~~
\frac{1}{\sqrt{Y_{\Sigma_1}^2+ Y_{\Sigma_2}^2}}
\pmatrix{{ {Y_{\Sigma_1}}} \cr {Y_{\Sigma_2}} \cr {Y_{\Sigma_3}}} \nn
\label{eq:lowvac}
\ee 
respectively. Clearly, two of the light neutrinos emerge as 
massless while the third one gets mass, which is in conflict with the low 
energy neutrino data. Similarly if one has only matrix $B$ which comes as
a consequence of R parity violation  then  one also would obtain only one 
light neutrino to be massive, in general determining only the largest atmospheric mass
scale \cite{lspdecay1,Chun:2002vp} . However the
simultaneous presence of the matrix $A$, $B$ and $C$ in Eq. (\ref{eq:mnuapprox})  
make a second eigenvalue  non-zero,
while the third one remains zero. With the choice $\tilde{u}$ as of the same
order of $u$, the third term  in Eq. (\ref{eq:mnuapprox}) would be suppressed
compared to the first two terms. Hence the simultaneous presence of the matrix $A$ and $B$ are 
very crucial to get both the solar and atmospheric mass  splitting and the allowed oscillation
parameters. Eigenvalues of the full $M_\nu$ given in 
Eq. (\ref{eq:mnuapprox}) are
\be
m_1=0,~~~
m_{2,3} \sim \frac{1}{2}[W \mp \sqrt{W^2-V}], \\ \nn
\ee
where
\be
W=\frac{v_2^2}{2M} \sum_{i}{Y_{\Sigma_i}^2}+ \frac{\mu \alpha}{2} 
\sum_{i}u_i^2+\frac{\tilde{u}v_1 \alpha}{\sqrt{2}}\sum_{i}u_iY_{\Sigma_i},
\ee
and 
\be
V&=&4(\frac{v_2^2\mu\alpha}{4M}-\frac{\tilde{u}^2v_1^2 \alpha^2}{8})
[Y_{\Sigma_1}^2(u_2^2+u_3^2)+ Y_{\Sigma_2}^2(u_3^2+u_1^2)+
Y_{\Sigma_3}^2(u_1^2+u_2^2) \\ \nn 
&&-2(u_1u_2Y_{\Sigma_1}Y_{\Sigma_2}+
u_1u_3Y_{\Sigma_1}Y_{\Sigma_3}+u_2u_3Y_{\Sigma_2}Y_{\Sigma_3})].
\ee
Similarly we have obtained approximate analytic expressions for the 
mixing matrix, however the expressions obtained are too complicated and 
hence we do not present them here. Instead we show in Tables
\ref{tab:param1} and Table \ref{tab:param2} an example set of model 
parameters $(M^{1,2},M,\mu,Y_{\Sigma_i},v_{1,2},\tilde{u},u_{i})$ 
which give the experimentally allowed mass-square differences
$\Delta m^2_{21}$ and $\Delta m^2_{31}$ and mixing angles $\sin^2
\theta_{12}$,  $\sin^2 \theta_{23}$ and $\sin^2 \theta_{13}$, as well as the
total neutrino mass $m_t$. The values obtained for these 
neutrino parameters for the model points given in  Tables
\ref{tab:param1} and Table \ref{tab:param2} is shown 
in Table \ref{tab:massmix}. We have presented these results 
assuming $\ma > 0$ (normal hierarchy).

\begin{table}[h]
\begin{center}
\begin{tabular}{|c|c|c|c|c|c|c|}
\hline
$M^1$ (GeV) & $M^2$ (GeV) & $M$ (GeV) & $\mu$ (GeV) & ${Y_{\Sigma_1}}$ & ${Y_{\Sigma_2}}$ 
& ${Y_{\Sigma_3}}$ \cr
\hline
$300$ & $600$ & $353.24$ & $88.31$ & $5.62\times 10^{-7}$ & $8.72\times 10^{-7}$ & $3.84 \times 10^{-8}$ \cr
\hline
\end{tabular}
\caption{\label{tab:param1}
Sample point in the parameter space  for the case of normal hierarchy. $M^{1,2}$ is the gaugino mass parameter, $\mu$ is higgsino mass parameter, $M$ is triplet fermion mass parameter, ${Y_{\Sigma_1}}$, ${Y_{\Sigma_2}}$ and ${Y_{\Sigma_3}}$  correspond  to the superpotential coupling between the standard model Lepton superfields $\hat{L_i}$, SU(2) triplet superfield  $\hat{\Sigma}_R^{0c}$ and Higgs superfield ${\it \hat{H_u}}$.
}
\end{center}
\end{table}

\begin{table}[h]
\begin{center}
\begin{tabular}{|c|c|c|c|c|c|}
\hline
$v_1$ (GeV) & $v_2$(GeV) & $\tilde{u}$(GeV) & $u_1$(GeV) & $u_2$ (GeV)& $u_3$(GeV) 
 \cr
\hline
$10.0$ & $100.0$ & $5.74 \times 10^{-3}$ & $1.69 \times 10^{-5}$ & $9.55 \times 10^{-5}$ & $1.26 \times 10^{-4}$  \cr
\hline
\end{tabular}
\caption{\label{tab:param2}Sample point in the parameter space  for the case of normal hierarchy. $v_{1,2}$ are the vacuum expectation values of $H_{d,u}^0$ fields respectively, $\langle {\nu_{L_i}}\rangle =u_i $ for i=1,2,3 and $\tilde{u}$ is the vacuum expectation value of triplet sneutrino state $\tilde{\Sigma}_R^{0c}$.
}
\end{center}
\end{table}
\begin{table}[h]
\begin{center}
\begin{tabular}{|c|c|c|c|c|c|}
\hline
$\Delta m^2_{21}$ ($eV^2$) & $\Delta m^2_{31}$ ($eV^2$) & $\sin^2 \theta_{12}$ & $\sin^2 \theta_{23}$ & $\sin^2 \theta_{13 }$
& $m_t$ ($eV$)
 \cr
\hline
$7.44 \times 10^{-5}$ $eV^2$ & $2.60 \times 10^{-3}$ $eV^2$ &  $0.33$ & $0.507$ & $4.34 \times 10^{-2}$ & $10^{-2}$ \cr
\hline
\end{tabular}
\caption{\label{tab:massmix}
Values for neutrino oscillation parameters for the input parameters specified in Table\ref{tab:param1} and in Table \ref{tab:param2}. 
}
\end{center}
\end{table}
 

At this point we would like to comment on the possibility of  radiatively-induced neutrino mass generation in our model. In a generic 
R-parity violating MSSM,  both the lepton number and baryon number violating operators are present. Apart from
the tree level bilinear term $\hat{L}\hat{H}_u$ which mixes neutrino with higgsino, the trilinear lepton
number violating operators $\lambda \hat{L}\hat{L}\hat{E}^c$, $\lambda^{\prime}\hat{L}\hat{Q}\hat{D}^c$ and also the bilinear 
operator  $\hat{L}\hat{H}_u$  contribute to the radiatively-induced neutrino mass generation \cite{susybneu1, oneloop, twoloop}. In general there 
could be several loops governed by the $\lambda \lambda$, $\lambda^{\prime}\lambda^{\prime}$, $BB$, $\epsilon B$ couplings.
However in the spontaneous  R-parity violating model, working in the weak basis we would only have  $\hat{L}\hat{H}_u$ operator  coming from $\hat{H}_u{\hat{\Sigma}_R}^{c}\hat{L}$ term in the superpotential. Similarly in the scalar potential we would obtain $H_u\tilde{L}$ coupling coming out from ${H}_u{\tilde{\Sigma}_R}^{c}\tilde{L}$ term in Eq. (\ref{eq:lsigmasoft}). Hence we can have loops governed by  $A_{\Sigma}A_{\Sigma}$ couplings, like the $BB$ loop in Fig. 3 of \cite{susybneu1}. For the sake of completeness we have presented the diagram in Fig. \ref{fig:AsigAsig}. The one loop corrected neutrino mass coming out from this diagram would be $m_{\nu} \sim \frac{g^2\tilde{u}^2}{64\pi^2 cos^2 \beta}
\frac{A_{\Sigma_i}A_{\Sigma_j}}{\tilde{m}^3} $. In general for moderate values of $\cos \beta$, if one chooses the average slepton mass $\tilde{m}$  to be in the TeV order and the soft supersymmetry breaking coupling $A_{\Sigma} \sim 10^2 $ GeV,  then because  $\tilde{u} \sim 10^{-3}$ GeV, the contribution coming from this diagram would be roughly suppressed by a factor of $10^{-2}$  compared to the tree level neutrino mass.  Similar  conclusion  can be drawn  for the $\epsilon A_{\Sigma}$ \cite{susybneu1} loop induced neutrino mass, shown in Fig. \ref{fig:epsilonAsig}. In our model the R parity violating  $\lambda $ and $\lambda^{\prime}$ couplings do not get generated in the weak basis. However, from the R-parity conserving superpotential given in Eq. (\ref{eq:wmssm}) it would be possible to generate $\lambda$ and $\lambda^{\prime}$ couplings via mixings and only after transforming to a mass basis. Hence, in general for our model we would expect the contributions coming from the 
$\lambda \lambda$ and $\lambda^{\prime}\lambda^{\prime}$ loops to be suppressed compared to the tree level neutrino masses. 
Apart from these different bilinear and trilinear radiatively induced neutrino mass generation,  there
could be another source of radiative neutrino mass, namely the non-universality in the slepton mass matrices. 
In the R parity conserving limit the analysis would be same as of \cite{Davidson:1998bi}, only triplet fermions in our model are generating the Majorana sneutrino masses $\tilde{\nu}_i\tilde{\nu}_j$. However, in this work we stick to the universality of the slepton masses, hence this kind of radiative neutrino mass generation will not play any non trivial  role. Due to the RG running from the high scale to the low scale the universal soft supersymmetry breaking slepton masses could possibly get some off-diagonal contribution  \cite{Casas:2001sr}, which we do not address in this present work.

\section{Collider Signature}
In this section we discuss very briefly about  the possibility of testing our
model  in collider experiments. Because R-parity is violated  in our model
there will be extra channels compared to the R-parity conserving minimal
supersymmetric standard model, which carry the  information about R-parity
violation. As  R-parity gets broken, the triplet neutral heavy lepton
$\Sigma_R^{0c}$ and standard model neutrino ${\nu_L}_i$  mix  with the neutral
higgsino $\tilde{H}_u^0$ and  $\tilde{H}_d^0$, with  bino $ \tilde{\lambda}^0$ and
wino  $\tilde{\lambda}^3$, in addition to the  usual R-parity conserving Dirac mixing
between them. Hence in the mass basis with one generation of heavy triplet
fermion, there will be 5 neutralinos in our model. We adopt the convention
where the neutralino state $\tilde{\chi}_i^0$ are arranged according to the
descending order of their mass and  $\tilde{\chi}_5^0$ is the lightest
neutralino. If one adopts gravity mediated supersymmetry breaking as the
origin of soft supersymmetry breaking Lagrangian, then the lightest neutralino
is in general  the lightest supersymmetric particle (LSP). But  as R-parity is
broken, in any case LSP will not  be stable\cite{lspdecay,lspdecay1}. 
For MSSM, the 
production mechanism of neutralinos and sneutrinos in colliders have been
extensively discussed in\cite{Rparity,neuprod}. 
Depending on the
parameters, the lightest neutralino can be gaugino dominated or higgsino
dominated. 
In our model in addition
to the standard model neutrinos we also have a heavy triplet neutral fermion
$\Sigma_R^{0c}$. Hence in this kind of model where  heavy triplet fermions are present, 
 the lightest neutralino can also be
$\Sigma_R^{0c}$ dominated. Detail analysis of the collider signatures of model
and implications for LHC will be presented in a forthcoming paper. Here we
present a qualitative discussion on the possible neutralino, sneutrino, 
slepton and chargino decay modes.

\begin{itemize}
\item (A) Neutralino two body decay: As R-parity is violated, the lightest
  neutralino can decay through R-parity violating decay modes. It can decay via the two body decay mode
  $\tilde{\chi}_5^0 \to lW$ or $\tilde{\chi}_5^0 \to \nu Z$. Other heavier
  neutralinos can decay to the lighter neutralino state  $\tilde{\chi}_i^0 \to
  \tilde{\chi}_j^0Z$, 
or through the decay modes $\tilde{\chi}_i^0 \to l^{\pm}W^{\mp} / \nu Z $. The
gauge boson can decay  leptonically or hadronically producing 
\be
\tilde{\chi}_i^0 &\to& l^{\pm} W^{\mp} \to l^{\pm} l^{\mp}+ \not{E_T}, \nn \\
\tilde{\chi}_i^0 &\to& l^{\pm} W^{\mp} \to l^{\pm}+2j,\nn \\
\tilde{\chi}_i^0  &\to& \nu Z  \to  l^{\pm}l^{\mp}+ \not{E_T}, \nn \\
\tilde{\chi}_i^0   &\to&\nu Z  \to 2b+  \not{E_T}. \nn
\ee

\item (B) Sneutrino and slepton decay:  Because of R-parity violation the
  slepton can decay to a charged lepton and a neutrino \cite{sleptonlsp}  via
  $\tilde{l} \to \nu l$. The sneutrino can also have the possible decay
  $\tilde{\nu} \to l^{+} \l^{-}$. Note that in the explicit $R_{{p}}$
  violating scenario, this interaction term between
  $\tilde{\nu}l^{\pm}l^{\mp}$ and $\tilde{l}\bar{l}\nu$ would have significant
  contribution from $\lambda LLE^C$.  Here $l^{\pm}$, $\nu$ and $\tilde{l}$
  all are in mass basis. In the spontaneous R-parity violating scenario these
  interactions are  possible only after basis redefinition. The different
  contributions to the above mentioned interaction terms will come from the
  MSSM R-parity conserving term $\hat{H}_d\hat{L}\hat{E}^c$ as well as from
  the kinetic terms of the different superfields after one goes to the mass
  basis. 

\item (C) Three body neutralino decay modes: 
The other possible decay modes for the 
neutralino are $\tilde{\chi}_i^0 \to \nu \tilde{\nu}$,  $ l^{\pm}
\tilde{l}^{\mp}$, $ \nu h$. If the lightest neutralino $\tilde{\chi}_5^0$ is
the lightest supersymmetric particle, then the slepton or sneutrino would be
virtual. The sneutrino or slepton can decay through the R-parity violating
decay modes. Hence neutralino can have three body final states  such as
$\tilde{\chi}_i^0 \to  \nu \tilde{\nu} \to \nu l^{\pm} l^{\mp}$,
$\tilde{\chi}_i^0 \to  l^{\pm} \tilde{l}^{\mp}  \to l^{\pm}\nu l^{\mp} $.

\end{itemize}

Our special interest is in the case  where the lightest  neutralino state is
significantly  triplet fermion $\Sigma^0$ dominated. Besides the Yukawa
interaction, the triplet fermion $\Sigma^{\pm} \rm{,} \Sigma^0$  have gauge
interactions. Hence they could be produced at a significant rate 
in a  proton proton
collider such as LHC through gauge interactions. The triplet fermions
$\Sigma^{\pm}$ and $\Sigma^0$ could be produced  via $pp \to W^{\pm}/Z \to
\Sigma^{\pm} \Sigma^0/ \Sigma^{\mp}$ \footnote{By $\Sigma^{\pm} \rm{,} \Sigma^0$ we mean chargino
  state $\tilde{\chi}^{\pm}$ or neutralino state $\tilde{\chi}^0$
  significantly dominated by $\Sigma^{\pm} $ and  $ \Sigma^{0}$ 
respectively.} channels  apart from the   Higgs mediated
channels. In fact the production cross section of these triplet fermions
should be more compared to the production cross section of the singlet
neutrino dominated neutralino states\cite{numssm}, as for the  former case the
triplet fermions have direct interactions with the gauge bosons.  The decay
channels for these triplet neutral fermions are as  $ \Sigma^0 \to \nu Z $,
$\Sigma^0 \to l^{\pm} W^{\mp}$, $\Sigma^0 \to \nu h$ and also  $\Sigma^0 \to
\nu \tilde{\nu}^*, l \tilde{l}^*$ followed by R-parity violating  
subsequent decays
of the slepton/sneutrino.

                         Apart from the neutralino sector in our model, the
                         charged higgsino and charged winos mix  with standard
                         model leptons  and heavy charged leptons
                         $\Sigma^{\pm}$. Hence, just like 
in  the neutralino sector, there
                         will also be R-parity violating chargino decay. The
                         chargino $\tilde{\chi_i}^{\pm}$ can decay into the
                         following states, 
$\tilde{\chi_i}^{\pm} \to l^{\pm}Z$,  $   \nu W  $, $\nu h^{\pm}$ and also
to  $\tilde{\chi_i}^{\pm} \to l \tilde{\nu} \to ll^{\pm}l^{\mp} $, $
\tilde{\chi_i}^{\pm} \to \nu \tilde{l} \to \nu \nu l$. Just like 
as for the neutralinos, depending on the parameters,
in this kind of model  with heavy extra triplet fermions the lightest chargino could be
as well $\Sigma^{\pm}$ dominated. Moreover, because of R-parity violation, there will 
be  mixing between the the slepton and charged Higgs and sneutrino and neutral
Higgs. The sneutrino state 
$\tilde{\Sigma}^{\pm}$ can have the decays $\tilde{\Sigma}^{\pm} \to \nu
l^{\pm}$, $\tilde{\nu} W^{\pm}$, $\tilde{l}^{\pm}Z$. Similarly,  
$\tilde{\Sigma^0}$ can decay into 
$\tilde{\Sigma}^{0}  \to
l^{+}l^{-}$. Apart from these above mentioned decay channels, some 
other possible decay channels are, $\tilde{\Sigma}^{+} \to \bar{d}u$, $
\tilde{\Sigma}^0 \to u\bar{u}$, 
$\Sigma^{+} \to \tilde{u}\bar{d}$. 
We reiterate that by 
$\Sigma^{\pm,0}$ and $\tilde{\Sigma}^{\pm,0}$, we mean triplet dominated
chargino/neutralino  or sfermionic states. Note that for 
the R-parity conserving
seesaw scenario, these decay modes will be totally absent, as there is
no mixing between the triplet/singlet  fermion  
and the higgsino/gauginos and
mixing between sfermions and Higgs bosons.


We would also like to comment about the possibility of  lepton flavor violation 
in our model. For the non-supersymmetric Type III seesaw,  
the reader can find detailed  study on  lepton flavor violation in 
\cite{lfv}. Embedding  the 
triplet fermions in a supersymmetric framework opens up many new diagrams which can contribute to
 lepton flavor violation, for example  $\mu \to e \gamma$. In general 
the sneutrino,   triplet sneutrino, different sleptons and the charginos or neutralinos 
can flow within the loop \cite{lfvall,Casas:2001sr,carlos1996du,Allanach:2003eb}. In our model the R-parity violating effect comes very selectively.
The main contribution comes via the bilinear R-parity violating terms which get  only generated   
spontaneously.  Hence as discussed before, we ignore  $\lambda$ and $\lambda^{\prime}$ dominated diagrams 
\cite{carlos1996du}
and we expect that the lepton flavor violating contribution would be mainly governed by the soft supersymmetry 
breaking off-diagonal slepton mass contribution. Since we stick to the universal soft supersymmetry  breaking slepton 
masses, only  RG running  might  generate the off diagonal supersymmetry  breaking slepton masses
 \cite{Casas:2001sr}. In the R parity conserving limit the soft supersymmetry breaking slepton masses get a 
contribution  $ m^2_L \propto (3m_0^2+A_0^2)({Y_{\Sigma}Y^T_{\Sigma}}) log(\frac{M_X}{M_{\Sigma}})$ \footnote{For simplicity we have taken $Y_{\Sigma}$ to be real.}. For $m_0$,$ A_0 \sim $TeV  and $Y_{\Sigma} \sim 10^{-6}$, $m^2_L \sim 10^{-6}$ $log(\frac{M_X}{M_{\Sigma}})$$ \rm{GeV}^2$. The branching ratio would be roughly  $\frac{\alpha^3}{G_F^2}\frac{|m^2_L|^2}{\tilde{m}^8}tan^2 \beta$.  However, because of extremely small Yukawa $Y_{\Sigma} \sim  10^{-6} -10^{-8}$  our model would not violate the bound coming from 
$\mu \to e \gamma$. 

Before concluding this section  we present a qualitative comparision between  our model and  other models having $SU(2)$ triplet. 
There have been several studies on another class of models containing $SU(2)$ triplet  and $Y=2$
chiral superfields \cite{type2}. The triplet fermions in our model 
share a very distinguishing feature compared to the Higgs triplet models because of 
hypercharge  $Y=0$. While in  a model with Higgs  triplet  and in its supersymmetrized  version  one would
 inevitably have a doubly charged Higgs and higgsino, the model with  triplet fermion and its supersymmetrized version offer only 
singly charged and charge neutral fermions and their scalar superpartners. Moreover  though in our model we consider R parity violation,
but the R parity violation comes in a very specific way. For example  in \cite{AristizabalSierra:2003ix} 
the authors have extended the superfield contents by adding $SU(2)$ triplet $Y=2$ chiral superfield and in addition 
to this they have incorporated the explicit bilinear R parity violation. In our model we have extended 
the particle contents by adding one $SU(2)$ triplets $Y=0$ chiral superfield and in our work we address the issue 
of spontaneous 
R parity violation.  
In fact these two different class of models offer very different  collider signatures. 
There have been several studies on the collider signatures of Higgs triplets. 
The readers can look  at \cite{collidertypeall, Perez2008ha}.  The different channels are 
$H^{--} \to l_il_j$, $H^{++} \to W^{-}W^{-}$, $H^{++} \to H^{+}W^{+}$ and so on. However,  for 
Higgs mass in the range of $M_{H^{++}}\sim 300$ GeV  the dileptonic channel is the most
dominant one \cite{Perez2008ha}. In \cite{AristizabalSierra:2003ix}  the authors have related
the neutrino mixing angle with the dileptonic decay branching ratio. On the other hand in the 
fermionic triplet models the important decay modes are $\Sigma^{\pm} \to l^{\pm}h$, 
$\Sigma^{\pm} \to W^{\pm}\nu$, $\Sigma^{0} \to \nu h$, $\Sigma^{0} \to Z \nu$ and so on. 
Again, for non supersymmetric version, detailed  analysis on the  collider signatures   
could be found in \cite{collidertype3, collidertypeall}. In \cite{type3us} the authors 
have addressed the Type-III seesaw in the context of an additional Higgs which offers 
very drastic difference in the triplet fermion and Higgs phenomenology. In this present work we have addressed 
what could be interesting collider signatures if one embeds the triplet fermion  into a supersymmetric framework
 while incorporating  R parity violation very selectively. The hypercharge $Y=0$ and $Y=2$ $SU(2)$ triplet models would
also differ in the different lepton flavor violating processes. Above we have presented a 
qualitative discussion on the lepton flavor violation  for $Y=0$ triplet. Very detailed study on the 
lepton flavor violating processes,  mediated  by the Higgs triplet fields have been done in 
\cite{Kakizaki:2003jk, Abada:2007ux}. 
In general the doubly charged Higgs can mediate the $\mu \to 3e$ processes at the tree level, while
for the process $\mu \to e \gamma$ the Higgs triplet fields would contribute via loops. Some detailed analysis
have been presented in \cite{Kakizaki:2003jk, Abada:2007ux}.

\section{Conclusion}

In this work we have explored the possibility of spontaneous R-parity
violation in the context of basic MSSM  gauge group. 
Since the R-parity violating terms also break lepton number, 
spontaneous R-parity violation could potentially generate 
the massless mode Majoron. We avoid the problem of the Majoron 
by introducing explicit breaking of lepton number in the 
R-parity conserved part of the superpotential. 
We do this by adding a SU(2) triplet $Y=0$ matter
chiral superfield $\hat{\Sigma}_R^c$ 
in our model. The gauge  invariant bilinear term in these
triplet superfields provides explicit lepton number violation in our
model. The superpartners of the standard model neutrino and the 
triplet  heavy
neutrino states acquire vacuum expectation values, thereby breaking R-parity
spontaneously. 

From the minimization condition of the scalar potential, we 
showed that in our model, the vacuum expectation values of the 
superpartner of the triplet heavy neutrino and 
the standard model neutrinos turn out
to be proportional to each other. 
For the supersymmetry breaking soft masses  in the TeV
range,  smallness of neutrino mass ($\sim eV$) constraints the  standard model sneutrino
vacuum expectation value 
$u \sim  10^{-3}$ GeV. Since for small $u$ the sneutrino vacuum expectation values $u$ and $\tilde{u}$ are proportional, hence in
the absence of any fine tuned cancellation between the different soft
parameters, one can expect  that 
$\tilde{u}$ will be of the same order as $u$, i.e
$10^{-3}$ GeV. We  have analyzed the neutralino-neutrino mass matrix
and  have shown that the R-parity violation can have significant effect in
determining the correct neutrino mass and mixing. 
Since neutrino experiments demand at least two massive neutrinos, we 
restrict ourselves to 
only on one generation of $\hat{\Sigma}_R^c$ superfield. While in
R-parity conserving Type-III susy seesaw with only one generation of
triplet neutrino state $\Sigma_R^{0c}$, two of the standard model neutrinos 
turn out to be massless. If one invokes R-Parity violation 
in addition then one among the massless states can be made massive. We showed that 
the main impact on the neutrino oscillation parameters comes from 
the gaugino
mass dominated seesaw term in the neutrino mass matrix. 
The correct mass splitting and mixing angles for the low
energy neutrino could be achieved. Alongwith the neutralino-neutrino
mixing, we have chargino-charged lepton mixing in our model. 
Because of the presence 
of the triplet fermionic states, we have heavy charged fermions 
in our model. The heavy fermionic states modify the chargino-charged 
lepton mass matrix. In particular, we can have charginos which are 
heavy triplet fermion dominated. We also have qualitatively discussed on the radiatively induced 
neutrino mass for our model.


Finally we discussed the neutralino, chargino , slepton and   sneutrino decay
decays and the possible 
collider signature of this model. Because of R-parity violation the
neutralino and chargino can decay through a number of  R-parity violating decay channels
alongwith the possible R-parity violating decays of sneutrinos and
sleptons. In the context of our model, we have listed few of these 
channels. Depending
on the parameters of the 
neutralino-neutrino mass matrix, the lightest neutralino
could be a gaugino dominated or could be higgsino dominated. Since  heavy triplet fermions 
are present, hence the lightest neutralino state in this kind of models
 could 
even  
be triplet neutrino dominated. Similar kind of conclusion would hold for
the 
chargino states as well. 
Unlike the $SU(2)_L \times U(1)_Y$ singlet neutrino states, 
the triplet neutrino/sneutrino states have direct interaction with the
standard model gauge bosons. Hence the production cross section  of triplet
neutrino dominated neutralino should be more than the singlet neutrino
dominated neutralino states\cite{numssm}. The production cross section of
triplet charged lepton dominated chargino should also be 
significantly higher. We 
expect multileptonic final states associated with jets or missing energy as a
collider signature of our model. Detailed collider signatures and 
effective cross-sections of model warrants a separate study, 
which is underway and will be reported elsewhere. In addition to this we
also have presented a qualitive discussion on the lepton flavor violation 
in our model and the relative comparision between different models having
$SU(2)$ triplets with hypercharge $Y=0$ and $Y=2$.

\vglue 0.8cm
\noindent
{\Large{\bf Acknowledgments}}\vglue 0.3cm
\noindent
The authors wish to express their deep gratitude to  Biswarup Mukhopadhyaya 
for detailed discussions and support
at every stage of this work.
They also wish to thank 
Amitava  Raychaudhuri and Ashoke Sen 
for very  useful discussions and valuable comments. 
M.M wishes to thank Borut Bajc,
Francesco Vissani, Rathin Adhikari  and Pradipta Ghosh  for  discussions
and correspondence. 
This work has been supported by the Neutrino Project 
under the XI Plan of Harish-Chandra Research Institute (HRI).  
The authors acknowledge the HRI 
cluster facilities for computation.

\vglue 1.0cm

\begin{center}
{\LARGE{\bf \underline {Appendix}}}
\end{center}

\begin{appendix}

\section{Soft supersymmetry breaking lagrangian of MSSM}
The soft supersymmetry  breaking Lagrangian of this model is given by,  
\be
L^{\rm{soft}}=L_{\rm{MSSM}}^{\rm{soft}}+L_{\Sigma}^{\rm{soft}},
\ee
where  $L_{\Sigma}^{\rm{soft}}$  has been written in Eq. (\ref{eq:lsigmasoft}) and the MSSM soft supersymmetry breaking lagrangian has the following form, 
\be
-\mathcal L_{\rm{MSSM}}^{\rm{soft}} &=&
(m_{\tilde{Q}}^2)^{ij} {\tilde Q^{\dagger}_i} \tilde{Q_j}
+(m_{\tilde u^c}^{2})^{ij}
{\tilde u^{c^*}_i} \tilde u^c_j
+(m_{\tilde d^c}^2)^{ij}{\tilde d^{c^*}_i}\tilde d^c_j
+(m_{\tilde{L}}^2)^{ij} {\tilde L^{\dagger}_i}\tilde{L_j} \nonumber \\
&+&(m_{\tilde e^c}^2)^{ij}{\tilde e^{c^*}_i}\tilde e^c_j 
+ m_{H_d}^2 {H^{\dagger}_d} H_d + m_{H_u}^2 {H^{\dagger}_u} H_u +(bH_uH_d+ \rm{H.c.}) \nn \\
&+&  \left[
-A_u^{ij} H_u\tilde Q_i \tilde u_j^c +
A_d^{ij} H_d \tilde Q_i \tilde d_j^c +
A_e^{ij} H_d \tilde L_i \tilde e_j^c + \text{H.c.}  \right] 
\nonumber \\
 &+&\frac{1}{2} \left(M^3 \tilde{g} \tilde{g}
+ M^2 \tilde{\lambda}^i \tilde{\lambda}^i + M^1 \tilde{\lambda}^0 
\tilde{\lambda}^0 + \text{H.c.} \right).
\label{Lsoft}
\ee
where $i$ and $j$ are generation indices, 
$m_{\tilde{Q}}^2$, $m_{\tilde{L}}^2$ and other terms in the first  two lines of the above equation represent the mass-square of different squarks, slepton, sneutrino\footnote {${{m^2_{\tilde{\nu}}}}$  represents the mass square of the superpartner of the standard model neutrino and  ${{m^2_{\tilde{\nu}}}}$=$m^2_{\tilde{L}} $.} and Higgs fields. In the third line the trilinear interaction terms have been written down and in the fourth line $M^3$, $M^2$ and  $M^1$ are respectively the  masses of the gluinos $\tilde{g}$, winos $\tilde{\lambda}^{1,2,3}$ and bino  $\tilde{\lambda}^0$. 

\section{Gaugino-lepton-slepton mixing}
In this section we write down explicitly all the interaction terms generated
from $W_{\Sigma}$, as well as the gaugino-triplet leptons-triplet sleptons
interaction terms originating  from $L_{\Sigma}^k$. As has already been discussed  in section 2, $W_{\Sigma}$ is given by
\be
W_{\Sigma}= -Y_{\Sigma_i}\hat{H}_u^T (i\sigma_2) \hat{{\Sigma}}_R^c \hat{L_i}+ \frac{M}{2} Tr[\hat{\Sigma}_R^c \hat{{\Sigma}}_R^c].
\label{eq:wsigma}
\ee
\be
W_{\Sigma}&=& \frac{{Y_{\Sigma_i}}}{\sqrt{2}} \hat{H}_u^0 \hat{\Sigma}_R^{0c} \hat{\nu}_{L_i} +{Y_{\Sigma_i}} \hat{H}_u^0\hat{\Sigma}_R^{-c} \hat{l_i}^-+\frac{{Y_{\Sigma_i}}}{\sqrt{2}} \hat{H}_u^+\hat{\Sigma}_R^{0c}\hat{l_i}^- -{Y_{\Sigma_i}} \hat{H}_u^+ \hat{\Sigma}_R^{+c}\hat{\nu}_{L_i} \nn \\
& & +\frac{M}{2} \hat{\Sigma}_R^{0c}\hat{\Sigma}_R^{0c} + M\hat{\Sigma}_R^{+c}\hat{\Sigma}_R^{-c}. 
\label{eq:wsexp}
\ee
In Table. \ref{tab:intwsigma} we show all the trilinear  interaction terms generated from $Y_{\Sigma_i}\hat{H}_u \hat{\Sigma}_R^c \hat{L_i}$.
\begin{table}[h]
\begin{center}
\begin{tabular}{|c|c|c|c|}
\hline
$\frac{Y_{\Sigma_i}}{\sqrt{2}} \hat{H}_u^0 \hat{\Sigma}_R^{0c}\hat{\nu}_{L_i} $ & $ Y_{\Sigma_i} \hat{H}_u^0\hat{\Sigma}_R^{-c} \hat{l_i}^- $ & $ \frac{Y_{\Sigma_i}}{\sqrt{2}} \hat{H}_u^+ \hat{\Sigma}_R^{0c}\hat{l_i}^-$ & $ -Y_{\Sigma_i} \hat{H}_u^+ \hat{\Sigma}_R^{+c}\hat{\nu}_{L_i}$ \cr
\hline

$-\frac{Y_{\Sigma_i}}{\sqrt2} \tilde{H}_u^0{\Sigma}_R^{0c} \tilde{\nu}_{L_i}$ & $-{Y_{\Sigma_i}} \tilde{H}_u^0 \Sigma_R^{-c} \tilde{l_i^{-}}$  & $-\frac{Y_{\Sigma_i}}{\sqrt2} {H_u^+}{\Sigma_R^{0c}} {l_i^{-}}$  & ${Y_{\Sigma_i}} \tilde{H}_u^+ \Sigma_R^{+c} {\nu_{L_i}}$  \cr
$-\frac{Y_{\Sigma_i}}{\sqrt2} \tilde{H}_u^0 \nu_{L_i} \tilde{\Sigma}_R^{0c} $ & $-{Y_{\Sigma_i}} \tilde{H}_u^0 \tilde{\Sigma}_R^{-c} {l_i^{-}}$  & $-\frac{Y_{\Sigma_i}}{\sqrt2} \tilde{H}_u^+ \Sigma_R^{0c} \tilde{l_i}^{-}$ & ${Y_{\Sigma_i}} \tilde{H}_u^+ \tilde{\Sigma}_R^{+c} {\nu_{L_i}}$  \cr
$-\frac{Y_{\Sigma_i}}{\sqrt2} {H_u^0} \Sigma_R^{0c}\nu_{L_i} $ & $-{Y_{\Sigma_i}} {H_u^0} \Sigma_R^{-c} {l_i^{-}}$   & $-\frac{Y_{\Sigma_i}}{\sqrt2} \tilde{H}_u^+ \tilde{\Sigma}_R^{0c} {l_i^{-}}$  & ${Y_{\Sigma_i}} \tilde{H}_u^+ \Sigma_R^{+c} \tilde{\nu}_{L_i}$  \cr
\hline
\end{tabular}
\caption{\label{tab:intwsigma}
Trilinear interaction terms  between standard model leptons/sleptons, Higgs/higgsinos and the SU(2) triplet fermions/sfermions. These interaction terms originate from the superpotential $W_{\Sigma}$. 
}
\end{center}
\end{table}

\noindent
The kinetic terms of the $\hat{\Sigma}_R^c$ field is given by 
\be 
L_{\Sigma}^{k}= \int d^4\theta   Tr[{\hat{\Sigma_R^c}}^{\dagger}e^{2gV}\hat{{\Sigma_R^c}}].
\label{eq:sigmak}
\ee
where $V$ represents the SU(2) vector supermultiplets.
From the above kinetic term one will get the following  gaugino-triplet fermion-triplet sfermion interactions 
\be
-{L_{\Sigma-\tilde{\Sigma}- \tilde{\lambda}^3}}= \frac{g}{\sqrt{2}} (({\tilde{\Sigma}_R^{-c}})^* \tilde{\lambda}^3 {\Sigma_R^{-c}}-({\tilde{\Sigma}_R^{+c}})^* \tilde{\lambda}^3 {\Sigma_R^{+c}})+h.c,
\label{eq:fsfg1}
\ee
\be
-{L_{\Sigma-\tilde{\Sigma}- \tilde{\lambda}^+}}=\frac{g}{\sqrt{2}} (( {\tilde{\Sigma}_R^{0c}})^* \tilde{\lambda}^+ {\Sigma_R^{+c}}-({\tilde{\Sigma}_R^{-c}})^* \tilde{\lambda}^+ {\Sigma_R^{0c}}) +h.c,
\label{eq:fsfg2}
\ee
\be
-{L_{\Sigma-\tilde{\Sigma}- \tilde{\lambda}^-}}=\frac{g}{\sqrt{2}}( ({\tilde{\Sigma}_R^{+c}})^* \tilde{\lambda}^- {\Sigma_R^{0c}}-({\tilde{\Sigma}_R^{0c}})^* \tilde{\lambda}^- {\Sigma_R^{-c}}) + \rm{h.c}.
\label{eq:fsfg3}
\ee
Note that the mixing terms between the gauginos and triplet fermions  $\tilde{\lambda}^+ {\Sigma_R^{+c}}$ and $\tilde{\lambda}^- {\Sigma_R^{-c}}$ contribute to the color singlet charged fermion mass matrix Eq. (\ref{eq:chmssm}) and these mixing terms are  generated from Eq. (\ref{eq:fsfg2}) and Eq. (\ref{eq:fsfg3}) respectively, once  $\tilde{\Sigma}_R^{0c}$  gets vacuum expectation value. In addition to these interaction terms  between gauginos, triplet leptons and triplet sleptons, we also explicitly write down  the gaugino-standard model lepton-slepton  interaction terms  which will be generated from the kinetic term of the $\hat{L_i}$ superfields
\be 
L_L^k= \int d^4\theta {\hat{L_i}^{\dagger}e^{2gV+2g^{\prime}V^{\prime}}\hat{L_i}}.
\label{eq:lkin}
\ee
\be
L_L^k &=& -\frac{g}{\sqrt{2}} {{\tilde{\nu}_{L_i}}}^*\tilde{\lambda}^3 {\nu_{L_i}} - \frac{g}{\sqrt{2}} {\tilde{l_i}}^*\tilde{\lambda}^3 {l_i}^--g{\tilde{\nu}_{L_i}}^*\tilde{\lambda}^+ {l_i}^--g\tilde{l_i}^* \tilde{\lambda}^{-}{\nu_L}_i \nn \\
&& + \frac{g^{\prime}}{\sqrt{2}} \tilde{\nu}_{L_i}^* \tilde{\lambda}^0 {\nu_{L_i}}+ \frac{g^{\prime}}{\sqrt{2}} {\tilde{l}_i}^* \tilde{\lambda}^0 {l_i^-}+ \rm{h.c}.
\label{eq:lkinexp}
\ee
Similar interaction terms would be generated from $\hat{E^c}$ kinetic term and kinetic terms of other superfields like the Higgs $\hat{H}_{u,d}$ and other quarks. Looking at Eq. (\ref{eq:lkinexp}) it is clear that once the sneutrino fields ${\tilde{\nu}_{L_i}}$  get vacuum expectation values  the first and fifth term of the above equation would contribute to gaugino-neutrino mixing while the third term would contribute in the chargino-charged lepton mixing, as have already  been shown in  Eq. (\ref{eq:massnm}) and in Eq. (\ref{eq:chmssm}).

\section{The Neutrino Mass, $Y_{\Sigma}$ and $u$}
In this section we discuss in detail how the smallness of neutrino mass  restricts  the order of magnitude of the Yukawa $Y_{\Sigma}$ and the sneutrino vacuum expectation value $u$.
Below we write the  analytical expression of the low energy neutrino mass of our model. The $3\times 3$ neutrino mass matrix is 
\be
M_{\nu} \sim - {m_D}^T{M^{\prime}}^{-1} m_D,
\ee
where $m_D$ and $M^{\prime}$ have already been given in Eq. (\ref{eq:mdt}) and  in Eq. (\ref{eq:masnm}) respectively. With the specified  $m_D$ and $M^{\prime}$, the $3 \times 3$ standard model  neutrino mass matrix $M_{\nu}$ has the following form,
\be
-M_{\nu}&=& 2 {v_2}^2 \alpha_1 A  + 2M \mu^2 \alpha_t B + \sqrt{2}\alpha_t \tilde{u}v_1M \mu C +M\alpha_t v_1^2 \tilde{u}^2 A\nn \\
& & -{\sqrt{2}\alpha_t \tilde{u}v_1^2v_2\sum_{i=1,2,3}{u_iY_{\Sigma_i}}} A-
{\alpha_tv_1v_2\mu} F.
\label{eq:mnuexact}
\ee
The matrix $A$, $B$ and $C$ have already been presented  in Eq. (\ref{eq:A}), Eq. (\ref{eq:B}) and Eq. (\ref{eq:C}). The  matrix F has the following form,
\be
F=\pmatrix{F_{11} & F_{12} & F_{13} \cr
F_{12} & F_{22} & F_{23} \cr
F_{13} & F_{23} & F_{33}},
\ee
where the  elements of F are 
\be 
F_{ij}(i \neq j)=Y_{\Sigma_i}Y_{\Sigma_j}(u_i^2+u_j^2)+u_iu_j(Y_{\Sigma_i}^2+Y_{\Sigma_j}^2)+u_kY_{\Sigma_k}(u_iY_{\Sigma_j}+u_jY_{\Sigma_i}),
\label{eq:fij}
\ee
and
\be
F_{ii}=2Y_{\Sigma_i}^2u_i^2+2u_iY_{\Sigma_i}\sum_{k \neq i}{Y_{\Sigma_k}u_k}.
\label{eq:fii}
\ee
The indices {\it i, j, k} in Eq. (\ref{eq:fij}) and the index {\it i} in Eq. (\ref{eq:fii}) are  {\it{not summed}} over. 
$\alpha_t$ has this following expression,
\be
\alpha_t= \frac{(M_1g^2+M_2{g^{\prime}}^2)}{4MM_1M_2\mu^2-4M\mu(M_1g^2+M_2{g^{\prime}}^2)v_1v_2-(M_1g^2+M_2{g^{\prime}}^2)v_1^2({\Sigma_i{u_iY_{\Sigma_i}}})^2} ,
\label{eq:alphatu}
\ee
while the parameter $\alpha_1$ is 
\be
\alpha_1=\frac{M_1M_2\mu^2-(M_1g^2+M_2{g^{\prime}}^2)v_1v_2\mu}{4MM_1M_2\mu^2-4M\mu(M_1g^2+M_2{g^{\prime}}^2)v_1v_2-(M_1g^2+M_2{g^{\prime}}^2)v_1^2({\Sigma_i{u_iY_{\Sigma_i}}})^2}
.
\label{eq:alpha1}
\ee
Below we show how the choice of TeV  scale gaugino masses and the triplet
fermion mass $M$ dictate the sneutrino vacuum expectation value $u$ to be
smaller than  $10^{-3}$ GeV and the Yukawa $Y_{\Sigma}\leq 10^{-5}$ to have
consistent small ( $\leq eV$) standard model neutrino mass. We consider the  following
three  cases and show that only  Case (A) is viable. 

\vspace*{0.2cm}
Case (A):  If one assumes  $ M M_1M_2 \mu^2 $ and $M \mu (M_1g^2+M_2{g^{\prime}}^2)v_1v_2 \gg
(M_1g^2+M_2{g^{\prime}}^2)u_i^2Y_{\Sigma_i}^2v_1^2 $,  the 
first and
second term dominate over the third 
one\footnote{Most likely between the first and 
second terms, the first
term would have larger value for the choice of TeV scale gaugino mass.},
in the
denominator of Eq. (\ref{eq:alphatu}) and  Eq. (\ref{eq:alpha1}). 
Then  $\alpha_t$ and $\alpha_1$ simplify to
\be
\alpha_t \sim \frac{(M_1g^2+M_2{g^{\prime}}^2)}{4MM_1M_2\mu^2}+...,
\ee
and
\be
\alpha_1 \sim \frac{1}{4M}+... .
\ee
The neutrino mass matrix Eq. (\ref{eq:mnuexact}) will have the following form,
\be
M_{\nu} &\sim& \frac{v_2^2 Y_{\Sigma}^2}{2M}+ \frac{u^2(M_1g^2+M_2{g^{\prime}}^2)}{2M_1M_2}+\sqrt{2}\tilde{u}v_1uY_{\Sigma}\frac{(M_1g^2+M_2{g^{\prime}}^2)}{4M_1M_2\mu}+ \frac{v_1^2 \tilde{u}^2Y_{\Sigma}^2(M_1g^2+M_2{g^{\prime}}^2)}{4M_1M_2\mu^2} \nn \\
&& -\sqrt{2}\tilde{u}v_1^2v_2uY_{\Sigma}^3\frac{(M_1g^2+M_2{g^{\prime}}^2)}{4MM_1M_2\mu^2}- v_1v_2\mu u^2Y_{\Sigma}^2\frac{(M_1g^2+M_2{g^{\prime}}^2)}{4MM_1M_2\mu^2}.
\label{eq:mnua}
\ee
The order  of $Y_{\Sigma}$ and $u$ would be determined  from the first 
two terms of the above equation  respectively. The 
fourth,  fifth and sixth terms of
Eq. (\ref{eq:mnua}) cannot determine  the order of $Y_{\Sigma}$ and $u$.  
To show this with an example  let us consider the contribution coming from the
fourth term $\frac{(M_1g^2+M_2{g^{\prime}}^2)v_1^2\tilde{u}^2Y_{\Sigma}^2}{M_1M_2
  \mu^2}$. Contribution of the order of 1 eV 
from this term demands $Y_{\Sigma}^2 \tilde{u}^2
\sim 10^{-4}  \rm{GeV}^2$ for $M^{1,2}$ in the TeV range, $\mu \sim 10^2$ GeV
  and $v_1$ in the GeV
range\footnote { As an example $v_1 \sim 10$ GeV.}. But if one assumes that
$Y_{\Sigma} \sim 1$ and $\tilde{u}^2 \sim 10^{-4}$$ GeV^2$ then we get 
the contribution from the first term of Eq. (\ref{eq:mnua}) $\frac{v_2^2Y_{\Sigma}^2}{M} \gg 1 \rm{eV}$, 
 for $M \sim $ TeV and  $v_2 \sim
100 $ GeV. The  choice $Y_{\Sigma}^2 \sim 10^{-4}$ and $\tilde{u} \sim 1$ GeV
also leads to $\frac{v_2^2Y_{\Sigma}^2}{M} \gg 1 \rm{eV}$ for $v_2 \sim 10^2$ GeV. The other option is
to have  $Y_{\Sigma}^2 \sim 10^{-12}$ and $\tilde{u}^2 \sim 10^8  \rm{GeV}^2$
so that $Y_{\Sigma}^2 \tilde{u}^2 \sim 10^{-4}  \rm{GeV}^2$. But to satisfy
this choice  one needs $10^{7}$ order of magnitude hierarchy  between the two
vacuum expectation values 
$\tilde{u}$ and $u$ \footnote{As the scale of $u$ is fixed from the second 
  term of Eq. (\ref{eq:mnua}) and $u <10^{-3}$ GeV.}.  
This in turn demands acute 
hierarchy between the different soft supersymmetry breaking mass terms
in Eq. (\ref{eq:uut}). Similarly,  $Y_{\Sigma}$ and $u$ could also not be
fixed from fifth and sixth terms of Eq. (\ref{eq:mnua}). Hence, we fix 
$Y_{\Sigma}$ and $u$ from the first two terms only.  The third term 
is larger than the fourth, fifth and sixth terms, but will still be
 subdominant compared to the first two terms \footnote{ As the third term $\propto \tilde{u}uY_{\Sigma}$
and from the 1st two terms $Y_{\Sigma}$ and $u$ will turn out to be small.} . 
The  contribution from the
first three terms  to the low energy neutrino mass matrix is 
\be
M_{\nu} \sim \frac{v_2^2 Y_{\Sigma}^2}{2M}+ 
\frac{u^2(M_1g^2+M_2{g^{\prime}}^2)}{2M_1M_2}+ 
\sqrt{2}\tilde{u}v_1uY_{\Sigma}\frac{(M_1g^2+M_2{g^{\prime}}^2)}{4M_1M_2\mu}.
\label{eq:mnuapp}
\ee
 It is straightforward to see from the first term of this approximate
 expression,   that $M_{\nu} \leq  1 eV$  demands $Y_{\Sigma}^2 \leq
 10^{-10}$ for  $M$ in the TeV range and $v_2 \sim 100$ GeV. On the other hand
 for $M_{1,2}$ also in the TeV range, the bound on sneutrino vacuum
 expectation value $u$ comes from the 
second term as $u^2 \leq 10^{-6}$ $GeV^2$. 
For the choice of $\tilde{u} \sim u$, which is a natural consequence of 
the scalar potential minimization conditions, one can check that the  third 
term in  Eq. (\ref{eq:mnuapp}) would be much smaller compared to the 
second term because of the presence of an additional suppression factor
$Y_{\Sigma}$. 
Hence, neutrino masses demand that 
$Y_{\Sigma} \leq 10^{-5}$
and $ u$ and $\tilde{u} \leq 10^{-3}$ GeV. One can explicitly check that 
for this above mentioned $Y_{\Sigma}$, $u$ and $\tilde{u}$ the contributions
coming from 4th, 5th and 6th terms of Eq. (\ref{eq:mnua}) are smaller compared
to the 1st three terms. 

\vspace*{0.3cm}

Case (B): If one assumes  $ M M_1M_2 \mu^2 \ll
(M_1g^2+M_2{g^{\prime}}^2)u_i^2Y_{\Sigma_i}^2v_1^2 $, which is possible to
achieve only for large value of the vacuum expectation value  $u$ and Yukawa
$Y_{\Sigma}$ \footnote{To satisfy this condition, the combination of
  $Y_{\Sigma}$ and $u$ has to be such that 
$u_i^2Y_{\Sigma_i}^2 \gg  10^{8} \rm{GeV^2}$
  for $M^{1,2},M \sim TeV$, $\mu$ in hundred GeV and $v_1$ in GeV range.}, so
that in the denominator of Eq. (\ref{eq:alpha1}) the 
third  term dominates over the
first two. Then the last term in Eq. (\ref{eq:mnuexact}) contributes as $ \sim
\frac{v_1v_2 \mu Y_{\Sigma}^2u^2}{v_1^2u^2Y_{\Sigma}^2} \sim \frac{\mu
  v_2}{v_1} \gg 1 \rm{eV}$ for $\mu$, $v_2$ and $v_1$ in the hundred GeV and
GeV range. Therefore, this limit is not a viable option.

\vspace*{0.3cm}

Case(C): We consider the last possibility  $ M M_1M_2 \mu^2 \sim
(M_1g^2+M_2{g^{\prime}}^2)u_i^2Y_{\Sigma_i}^2v_1^2 $, which also demands large
Yukawas and large vacuum expectation value $u$. If one considers  $M^{1,2}$ and
$M$ in the TeV range and $\mu$, $v_2$  in the hundred GeV range, then
demanding that the first term  in Eq. (\ref{eq:mnuexact}) 
should be smaller than eV 
results in the bound $Y_{\Sigma}^2 \leq 10^{-10}$. Hence, in order to
satisfy $ M M_1M_2 \mu^2 \sim
(M_1g^2+M_2{g^{\prime}}^2)u_i^2Y_{\Sigma_i}^2v_1^2 $, one needs $u \geq  10^9$ GeV
for $v_1$ in the GeV range.  For such large values of $u$, 
the second term in
Eq. (\ref{eq:mnuexact}) will give a very large contribution to 
the neutrino mass. This case is therefore also ruled out by  
the neutrino data.

Hence Case (B) and Case (C) which demand 
large $u$ and $Y_{\Sigma}$  are clearly inconsistent with the neutrino
mass scale,  and the only allowed possibility is Case (A). 
This case requires $u
\leq  10^{-3}$ GeV and $Y_{\Sigma} \leq 10^{-5}$, for the gaugino and triplet
fermion masses  $M^{1,2}$ and $M$ in the TeV range. As for small $u$,
$\tilde{u}$ and $u$ are proportional to each other, 
one obtains $\tilde{u}
\sim  10^{-3}$ GeV as well.


\end{appendix}

\newpage

\begin{figure}[tb]
\unitlength1mm
\SetScale{2.8}
\begin{boldmath}
\begin{center}
\begin{picture}(60,20)(0,0)
\Line(0,0)(15,0)
\Line(45,0)(15,0)
\Line(60,0)(45,0)
\DashCArc(30,0)(15,0,180){1}
\Text(13,6)[c]{$\tilde\nu_i$}
\Text(20,11)[c]{$\bullet$}
\Text(16,14)[c]{$A_{\Sigma_i}$}
\Text(47,5.6)[c]{$\tilde\nu_j$}
\Text(40,11)[c]{$\bullet$}
\Text(43,13.6)[c]{$A_{\Sigma_j}$}
\Text(-2,0)[r]{$\nu_i$}
\Text(62,-0.4)[l]{$\nu_j$}
\Text(30,18)[c]{$h,H,A$}
\Text(30,0)[c]{$\times$}
\Text(30,-5)[c]{$\chi_{\alpha}$}
\end{picture}
\end{center}
\end{boldmath}
\caption[a]{The $A_{\Sigma}A_{\Sigma}$ loop-generated neutrino mass.
The cross on the internal solid line represents the majorana mass for the neutralino and the blob 
on the dashed line represents the mixing between sneutrino and the neutral Higgs. }
\label{fig:AsigAsig}
\end{figure}
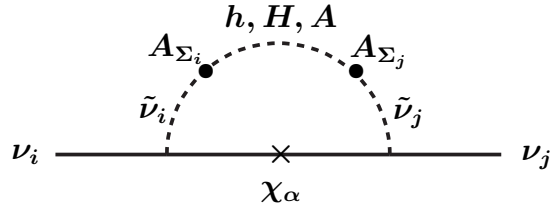
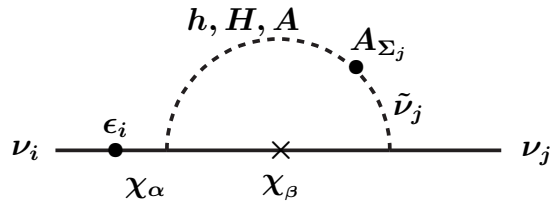
\begin{figure}[tb]
\unitlength1mm
\SetScale{2.8}
\begin{boldmath}
\begin{center}
\begin{picture}(60,20)(0,0)
\Line(0,0)(15,0)
\Line(45,0)(15,0)
\Line(60,0)(45,0)
\DashCArc(30,0)(15,0,180){1}
\Text(8,0)[c]{$\bullet$}
\Text(8,3)[c]{$\epsilon_i$}
\Text(47,5.6)[c]{$\tilde\nu_j$}
\Text(40,11)[c]{$\bullet$}
\Text(43,14)[c]{$A_{\Sigma_j}$}
\Text(-2,0)[r]{$\nu_i$}
\Text(62,-0.4)[l]{$\nu_j$}
\Text(25,17)[c]{$h,H,A$}
\Text(30,0)[c]{$\times$}
\Text(12,-5)[c]{$\chi_\alpha$}
\Text(30,-5)[c]{$\chi_{_\beta}$}
\end{picture}
\end{center}
\end{boldmath}
\caption[a]{
Neutrino Majorana mass generated by $\epsilon A_{\Sigma}$ loop. 
}
\label{fig:epsilonAsig}
\end{figure}
\end{document}